\documentclass[prd,twocolumn,showpacs,amsmath,amssymb]{revtex4}
\usepackage{graphicx}
\usepackage{dcolumn}
\usepackage{bm}

\newcommand{\Dp}{D^{\prime}}

\newcommand{\ag}{a_{3g}}
\newcommand{\aga}{a_{\gamma}}
\newcommand{\ac}{a_c}

\newcommand{\pip}{\pi^+}
\newcommand{\pim}{\pi^-}
\newcommand{\piz}{\pi^0}
\newcommand{\kap}{K^+}
\newcommand{\kam}{K^-}
\newcommand{\kaz}{K^0}
\newcommand{\kazb}{\overline{K}^0}
\newcommand{\kst}{K^*}
\newcommand{\kstp}{K^{*+}}
\newcommand{\kstm}{K^{*-}}
\newcommand{\kstz}{K^{*0}}
\newcommand{\kstzb}{\overline{K}^{*0}}

\newcommand{\rop}{\rho^+}
\newcommand{\rom}{\rho^-}
\newcommand{\roz}{\rho^0}

\newcommand{\etap}{\eta^{\prime}}

\newcommand{\psp}{\psi^{\prime}}

\newcommand{\jpsi}{J/\psi}

\newcommand{\EE}{e^+e^-}

\newcommand{\GG}{\gamma\gamma}

\newcommand{\kk}{K^+K^-}
\newcommand{\kskl}{K^0_SK^0_L}

\newcommand{\rpi}{\rho\pi}

\newcommand{\RP}{\rho\pi}

\newcommand{\rhopi}{\rho\pi}

\newcommand{\Heff}{{\cal H}_{eff}}
\newcommand{\Hz}{H_{0}}

\newcommand{\gz}{g_{0}}

\newcommand{\ra}{\rightarrow}
\newcommand{\beq}{\begin{equation}}
\newcommand{\eeq}{\end{equation}}
\newcommand{\beqn}{\begin{eqnarray}}
\newcommand{\eeqn}{\end{eqnarray}}
\newcommand{\beqns}{\begin{eqnarray*}}
\newcommand{\eeqns}{\end{eqnarray*}}
\def\eref#1{(\ref{#1})}
\def\Journal#1#2#3#4{{#1} {\bf #2}, #3 (#4)}

\def\NPA{Nucl. Phys. A}
\def\NPB{Nucl. Phys. B}
\def\PL{Phys. Lett. }
\def\PLB{Phys. Lett. B}

\def\PRL{Phys. Rev. Lett.}
\def\PRD{Phys. Rev. D}

\def\PRP{Phys. Rep.}

\def\PTPS{Prog. Theor. Phys. Suppl.}

\def\HEPNP{HEP \& NP}

\def\prd#1#2#3 {{~Phys. Rev. D {#1}, #2 (#3) }}  
\def\plb#1#2#3 {{~Phys. Lett. B {#1}, #2 (#3) }}  

\begin{document}

\title{Symmetry analysis involving meson mixing for charmonium decay}

\author{X.H.Mo$^{1,2}$
\\  \vspace{0.2cm} {\it
$^{1}$ Institute of High Energy Physics, CAS, Beijing 100049, China\\
$^{2}$ University of Chinese Academy of Sciences, Beijing 100049, China\\
}
}
\email{moxh@ihep.ac.cn}
\date{\today}
	
\begin{abstract}
In the light of $SU(3)$ flavor symmetry, the effective interaction Hamiltonian in tensor form is obtained by virtue of group representation theory. The strong and electromagnetic breaking effects are treated as a spurion octet so that the flavor singlet principle can be utilized as the criterion to determine the form of effective Hamiltonian for all charmonium two body decays. Moreover, a synthetical nonet is introduced to include both octet and singlet representations for meson description, and resorting to the mixing angle the pure octet and singlet states are combined into the observable pseudoscalar and vector particles, so that the empirically effective Hamiltonian can be obtained in a concise way. As an application, by virtue of this scenario the relative phase between the strong and electromagnetic amplitudes is studied for vector-pseudoscalar meson final state. In data analysis of samples taken in $\EE$ collider, the details of experimental effects, such as energy spread and initial state radiative correction are taken into consideration in order to make full use of experimental information and acquire the accurate and delicate results.
\end{abstract}
\maketitle

\section{Introduction}
Since the upgraded Beijing Electron-Positron Collider (BEPCII) and spectrometer detector (BESIII) started data taking in 2008~\cite{bes,yellow}, the colossal charm and charmonium data samples in the world were collected, especially the data at $\jpsi$ and $\psp$ resonance peaks, which provide an unprecedented opportunity to acquire useful information for understanding the interaction dynamics by analyzing various decay final states.

Although the Standard Model (SM) has been accepted as a universally appreciated theory basis in the high energy community, it is still hard to calculate the wanted experimental observable from the first principle of the SM for a great many of processes, especially when the strong interaction is involved.
Quantum chromodynamics (QCD) as a widely appreciated theory of strong interaction, has been proved to be very successful at high energy when the calculation can be executed perturbatively. Nevertheless, its validity at the nonperturbative regime needs more experimental guidance. As an exploratory step, it is necessary to develop a reliable and extensively applicable phenomenological model (PM).

The advantage of PM lies in that a well-defined PM contains few experimentally determined parameters which have clear physical meaning; moreover, with only few parameters determined from experiment, PM could
produce concrete results which can be directly confirmed or falsified by experiment and may guide further experimental searches. Such a model has a good relation with elementary principle of the theory, and if applicable, can be used for further theoretical refinement. This point is noteworthy for the time being, since the general QCD can hardly provide solutions for special problems; conversely, we have to
establish certain effective empirical model to advance our understanding for generic QCD principle.

As a matter of fact, many models are constructed~\cite{Kowalski:1976mc,Clavelli:1983,Haber,Seiden88,nMorisita91,rBaldini98,zmy2015,Baldini19,moxh2022,moxh2024}, the parametrization of various decay modes are obtained, such as the pseudoscalar and pseudoscalar mesons (PP), vector and pseudoscalar mesons (VP), octet baryon-pair, and so on. Especially in Ref.~\cite{moxh2024}, by virtue of $SU(3)$ flavor symmetry, the effective interaction Hamiltonian in tensor form is obtained according to group representation theory. In the light of flavor singlet principle, systematical parametrization is realized for all charmonium two-body decay, including both baryonic and mesonic final states. The parametrizations of $\psp$ or $\jpsi$ decaying to octet baryon pair, decuplet baryon pair, decuplet-octet baryon final state, vector-pseudoscalar meson final state, and pseudoscalar-pseudoscalar meson final state, are presented. However, in the previous study~\cite{moxh2024} mesons are merely treated as pure octet states while the actual particles are the mixing of both octet and singlet states. Therefore, this paper concentrates on the meson mixing issue. With nonet concept and mixing angle, both octet and singlet states are mixed into mass eigenvalue states so that the pragmatically effective Hamiltonian can be constructed in a concise way. Such a treatment finalizes the parametrization scheme proposed in the previous symmetry analysis of charmonium decay.

In the next section, the parametrization scheme will be expounded first, a synthetical nonet is introduced to combine both octet and singlet representations for meson description, then the effective Hamiltonian is obtained consecutively. The section that follows discusses in detail the experimental character of $\EE$ collider, then the successive section focuses on concrete data analysis for VP final states. A special section is used to discuss another fit method and some open questions on understanding fit results. After that is a summary section. Relegated to the appendix are two kinds of materials which are mainly concerned on calculation details.

\section{Analysis framework}\label{xct_alsfrk}
In $\EE$ collider experiment, the initial state is obviously flavorless, then the final state must be flavor singlet. Moreover, only the Okubo-Zweig-Iizuka (OZI) rule suppressed processes are considered and the final states merely involve light quarks, that is $u, d, s$ quarks. Therefore, $SU(3)$ group is employed for symmetry analysis. The key rule herein is the so called ``flavor singlet principle'' that determines what kinds of terms are permitted in the effective interaction Hamiltonian. Resorting to the perturbation language, the Hamiltonian is written as
\beq
\Heff = H_0 + \Delta H~,
\label{perturbaionhmtn}
\eeq
where $H_0$ is the symmetry conserved term and $\Delta H$ the symmetry breaking term, which is generally small compare to $H_0$. Since we focus on two-body decay, merely two multiplets, say ${\mathbf n}$ and ${\mathbf m}$, need to be considered. In the light of group representation theory, the product of two multiplets can be decomposed into a series of irreducible representations, that is
\beq
{\mathbf n} \otimes {\mathbf m} = {\mathbf l_1} \oplus {\mathbf l_2} \oplus \cdots \oplus {\mathbf
l_k}~.
\label{dcpsmoftwomtplt}
\eeq
The singlet principle requires that among the ${\mathbf l_j} (j=1, \cdots, k)$, only the singlet term, i.e., ${\mathbf l_j}={\mathbf 1}$ for certain $j$, is allowed in the Hamiltonian. Since this term is obviously $SU(3)$ invariant, it is called the symmetry conserved term, i.e., $H_0$.

Now turn to the $SU(3)$-breaking term. Following the recipe of the proceeding study~\cite{Muraskin1963,Gupta1964a,Gupta1964b}, the $SU(3)$-breaking effect is treated as a ``spurion'' octet. With this notion, in order to pin down the breaking term in the Hamiltonian, the products of this spurion octet with the irreducible representations ${\mathbf l_j} (j=1, \cdots, k)$ will be scrutinized, only the singlet term in the decomposition is valid in the Hamiltonian. Concretely,
\beq
{\mathbf l_j} \otimes {\mathbf 8} = {\mathbf q_1} \oplus {\mathbf q_2} \oplus \cdots \oplus {\mathbf
q_k}~,
\label{dcpsmofoctandirrds}
\eeq
then if and only if ${\mathbf q_i=1}$, the corresponding term is allowed in the Hamiltonian. Since such a kind of term violates $SU(3)$ invariance, it is called the symmetry breaking term. In a word, with the singlet principle, the effective interaction Hamiltonian can be determined unequivocally.

Generally speaking, the Hamiltonian term should be written as $\psi M_1 M_2$, where $\psi$ indicates the charmonium state and $M_1$ and $M_2$ are two multiplet components. Since $\psi$ is the same for the whole final state, and the relative strength of multiplet final state really matters, $\psi$ dependence is left implicit henceforth.

The tensor denotation is adopted in order to express all kinds of multiplets consistently. A tensor of rank $(u,v)$ reads
$$T^{i_1 i_2 \cdots i_u}_{j_1 j_2 \cdots j_v}~~,$$
or denoted as $T(u,v)$. The irreducible tensor of $SU(3)$ representation is traceless and totally symmetric in indices of the same type. The number of independent components in a multiplet is called its dimension, which is calculated by the formula
\beq
d\{T(u,v)\} = \frac{1}{2} (u+1)(v+1)(u+v+2)~.
\label{dmnoftsr}
\eeq
The baryon and meson of a multiplet can be expressed by the corresponding  components of the irreducible tensor, the characteristics of which can be used to calculate the eigenvalues of particle charge ($Q$), hypercharge ($Y$), and the third component of isospin ($I_3$), that is
\beq
\begin{array}{lcr}
Q   & = u_1-v_1-\frac{1}{3} (u-v)~~,\\
Y   & = -u_3+v_3+\frac{1}{3} (u-v)~~,\\
I_3 & = \frac{1}{2}(u_1-u_2)-\frac{1}{2} (v_1-v_2)~~,\\
\end{array}
\label{eqnforqyi3}
\eeq
where $u_i$ denotes the number of upper indices with value $i$, and  $v_j$ denotes the number of lower indices with value $j$. For $SU(3)$ group, $i,j=1,2,3$ and $u=u_1+u_2+u_3$ and $v=v_1+v_2+v_3$. The values $(Q,Y,I_3)$ of tensor component  indicate the corresponding physical particle, as examples, some commonly used octets of baryon and meson are displayed as follows~\cite{quangpham,lichtenberg,aHosaka} :
\beq
{\mathbf B}=
\left(\begin{array}{ccc}
\Sigma^0/\sqrt{2}+\Lambda/\sqrt{6} & \Sigma^+   & p    \\
\Sigma^-  & -\Sigma^0/\sqrt{2}+\Lambda/\sqrt{6} & n    \\
\Xi^-     & \Xi^0            & -2\Lambda/\sqrt{6}
\end{array}\right)~~,
\label{oktbyn}
\eeq
\beq
\overline{\mathbf B}=
\left(\begin{array}{ccc}
\overline{\Sigma}^0/\sqrt{2}+\overline{\Lambda/}\sqrt{6}
                    & \overline{\Sigma}^+ & \overline{\Xi}^+    \\
\overline{\Sigma}^- & -\overline{\Sigma}^0/\sqrt{2}+\overline{\Lambda/}\sqrt{6}
                                          & \overline{\Xi}^0    \\
\overline{p}        & \overline{n} & -2\overline{\Lambda}/\sqrt{6}
\end{array}\right)~~,
\label{atoktbyn}
\eeq
\beq
{\mathbf V}=
\left(\begin{array}{ccc}
\roz/\sqrt{2}+\omega/\sqrt{6} & \rop         & \kstp    \\
\rom        & -\roz/\sqrt{2}+\omega/\sqrt{6} & \kstz    \\
\kstm       & \kstzb                         &  -2 \omega/\sqrt{6}
\end{array}\right)~~,
\label{oktvtmsn}
\eeq
and
\beq
{\mathbf P}=
\left(\begin{array}{ccc}
\piz/\sqrt{2}+\eta/\sqrt{6} & \pip         & \kap    \\
\pim        & -\piz/\sqrt{2}+\eta/\sqrt{6} & \kaz    \\
\kam        & \kazb                        & -2 \eta/\sqrt{6}
\end{array}\right)~~.
\label{oktpsmsn}
\eeq
The corresponding tensor notations are respectively $B^i_j$, $\overline{B}^i_j$, $V^i_j$, and $P^i_j$, where the superscript denotes the row index of matrix and the subscript the column index.

The $SU(3)$-breaking effect is treated as a ``spurion'' octet, its additive quantum numbers, such as $Q,Y,I_3$, are set to be 0 while the multiplicative quantum numbers are set to be 1. Specially speaking, there are two kinds of spurion octet, one is denoted as ${\mathbf S}_{m}$ that indicates the strong breaking effect and is $I$-spin conserved breaking, the other is denoted as ${\mathbf S}_{e}$ that indicates the electromagnetic breaking effect and is $U$-spin conserved breaking. The tensor expressions for these two special octets read
\beq
({\mathbf S}_{m})^{i}_{j} = g_m \delta^i_3 \delta^3_j~,
\label{smdefn}
\eeq
and
\beq
({\mathbf S}_{e})^{i}_{j} = g_e \delta^i_1 \delta^1_j~.
\label{sedefn}
\eeq

It is noted that if $u$ and $v$ are switched in Eq.~\eref{eqnforqyi3}, the values of $Q,Y,I_3$ just change their signs, which means a particle turns into its antiparticle. If the charge conjugate operator is denoted as $\hat{C}$, then
$$ \hat{C}~ T(u,v) \to T(v,u) ~.$$
This can be confirmed for ${\mathbf V}$ and ${\mathbf P}$. But for baryon, since the particle and its antiparticle are in different multiplet, so the charge conjugate not only changes the signs of values $(Q,Y,I_3)$ but also keeps the mass of particle unchanged. According to this requirement, it yields
$$ \hat{C}~ T(u,v) \to \overline{T}(v,u) ~,$$
which can be confirmed for ${\mathbf B}$ and $\overline{\mathbf B}$.

Since the effective Hamiltonian must be $\hat{C}$-parity conserved, which means it should be unchanged under $\hat{C}$-parity transformation. This will constrain the form of the effective Hamiltonian. Let's take the baryon octet as an example. According to group theory, the product of two octets can be reduced as follows
\beq
{\mathbf 8} \otimes {\mathbf 8} = {\mathbf 1} \oplus {\mathbf 8} \oplus {\mathbf 8}
\oplus {\mathbf 10} \oplus {\mathbf 10^*} \oplus {\mathbf 27}~.
\label{twoeightrdn}
\eeq
The singlet (denoted as $[\overline{B} B]_0$) building from two octets reads
\beq
[\overline{B} B]_0 = \overline{B}^i_j B^j_i~~,
\eeq
Here Einstein summation convention is adopted, that is the repeated suffix, once as a subscript and once as a superscript, implies the summation. Under $\hat{C}$-parity transformation, $\hat{C}~[\overline{B} B]_0 \to [B \overline{B} ]_0 $. In order to keep $\hat{C}$-parity, the effective Hamiltonian should take the form  \beq
\Hz = \gz \cdot ([\overline{B} B]_0  + [B \overline{B} ]_0)~~.
\eeq
Next, by virtue of Eq.~\eref{twoeightrdn} there are two types of octet: an antisymmetric, or $f$-type, and  a symmetric, or $d$-type, which are respectively constructed as
\beq
([\overline{B} B]_f )^i_j = \overline{B}^i_k B^k_j -\overline{B}^k_j B^i_k~~,
\eeq
and
\beq
([\overline{B} B]_d )^i_j = \overline{B}^i_k B^k_j +\overline{B}^k_j B^i_k
-\frac{2}{3} \delta^i_j \cdot \overline{B}^i_j B^j_i~~.
\eeq
Under $\hat{C}$-parity transformation, $\hat{C}~([\overline{B} B]_d)^i_j \to - ([B \overline{B} ]_d)^j_i $ and $\hat{C}~([\overline{B} B]_f)^i_j \to ([B \overline{B} ]_f)^j_i $. Here the eigenvalues $\xi_{d,f}$ with values $\xi_{d}=+1$ and $\xi_{f}=-1$ (obviously $\xi_{0}=+1$) can be introduced to describe such effects. So in the effective Hamiltonian, two terms involved $[\overline{B} B]_f$ and $[\overline{B} B]_d$ should have the forms
$$[\overline{B} B]_{d,f} + \xi_{d,f}[B \overline{B} ]_{d,f}~~. $$

Nevertheless, the ${C}$-parity does not change the cross section of certain process, the terms with and without ${C}$-parity transformation furnish the same measurement results, as far as the relative strength is concerned, it proves expedient to keep only one kind of term. Therefore, the effective interaction Hamiltonian for octet-octet final state reads
\beq
\left.\begin{array}{rl}
\Heff = & \gz \cdot [\overline{B} B]_0    \\
        & + g_m \cdot ([\overline{B} B]_f )^3_3  + g_m^{\prime} \cdot ([\overline{B} B]_d )^3_3   \\
        & + g_e \cdot ([\overline{B} B]_f )^1_1  + g_e^{\prime} \cdot ([\overline{B} B]_d )^1_1 ~~~.
\end{array}\right.~~
\label{effhmtctp}
\eeq
The other results for decuplet-decuplet and decuplet-octet final states can be obtained similarly, the detailed information can be referred to Refs.~\cite{moxh2022,moxh2024}, together with the corresponding parametrization forms.

The effective Hamiltonian for meson final state can be obtained in the similar way, say, for ${\mathbf V} {\mathbf P}$, it reads
\beq
\left.\begin{array}{rl}
\Heff^{VP} = & \gz \cdot [V P]_0    \\
        & + g_m \cdot ([V P]_f )^3_3  + g_m^{\prime} \cdot ([V P]_d )^3_3   \\
        & + g_e \cdot ([V P]_f )^1_1  + g_e^{\prime} \cdot ([V P]_d )^1_1 ~~~.
\end{array}\right.~~
\label{effhmtvp0}
\eeq
However, unlike baryon, the particle and antiparticle are contained in the same meson multiplet, so $\hat{C}~{\mathbf V}/{\mathbf P} \to {\mathbf V}/{\mathbf P}$, which like the neutral particle such as $\piz$ with $\hat{C}~\piz \to \piz$. It is known that $\piz$ has the inherent ${C}$-parity, that is $\hat{C}~\piz= \eta_{\piz} \piz$, with $\eta_{\piz}=+1$. Herein, the generalized inherent ${C}$-parity for multiplet is introduced, and its value is set to be equal to that of the neutral particle in the multiplet. That is to say,
\beq
\left.\begin{array}{rl}
\hat{C}~{\mathbf P} = & \eta_{P} {\mathbf P}, ~\mbox{with}~\eta_{P}=\eta_{\piz}=+1~~;  \\
\hat{C}~{\mathbf V} = & \eta_{V} {\mathbf V}, ~\mbox{with}~\eta_{V}=\eta_{\roz}=-1~~.
\end{array}\right.~~
\label{cpvalofvap}
\eeq
With these convention, it has
\beq
\left.\begin{array}{rl}
\hat{C}~[V P]_f = & \eta_{V} \eta_{P} \xi_f [V P]_f~~,  \\
\hat{C}~[V P]_d = & \eta_{V} \eta_{P} \xi_d [V P]_d~~.
\end{array}\right.~~
\label{cpforvpdaf}
\eeq
The invariant of the effective Hamiltonian under ${C}$-parity transformation requires that $\eta_{V} \eta_{P} \xi_f $ or $\eta_{V} \eta_{P} \xi_d$ must be equal to $\eta_{\jpsi,\psp}$. Since $\eta_{\jpsi,\psp}=-1$, only $\eta_{\jpsi,\psp}=\eta_{V} \eta_{P} \xi_d$ is allowed, which means only the term $[V P]_d$ can exist in the effective Hamiltonian of ${\mathbf V} {\mathbf P}$ final state, that is
\beq
\Heff^{VP} = \gz \cdot [V P]_0
        + g_m \cdot ([V P]_d )^3_3  + g_e \cdot ([V P]_d )^1_1 ~.
\label{effhmtvpmsn}
\eeq
By virtue of the same criterion, the effective Hamiltonian of ${\mathbf P} {\mathbf P}$ final state reads
\beq
\Heff^{PP} =  g_m \cdot ([P P]_f )^3_3  + g_e \cdot ([P P]_f )^1_1 ~.
\label{effhmtppmsn}
\eeq
With the components given in Eqs.~\eref{oktvtmsn} and \eref{oktpsmsn}, the corresponding parametrization can be obtained and summarized in Tables~\ref{vpmsnform} and \ref{ppmsnform}, respectively. For the other octet final states, the parametrization scheme can be referred to the appendix.

\begin{table}[hbt]
\caption{\label{vpmsnform} Amplitude parametrization form for
decays of the $\psp$ or $\jpsi$ into $V P$ final states. General expressions in terms of singlet $A$ (by definition $A=g_0$), as well as charge-breaking ($D=g_e/3$) and mass-breaking terms ($\Dp=g_m/3$). The table can also be used for more similar decays by appropriate change in labeling, refer to appendix for more details. } \center
\begin{tabular}{lc}\hline \hline
  Final state    & Amplitude parametrization form  \\ \hline
  $\rho^{\pm}\pi^{\mp}$, $\roz \piz$  & $A+D-2\Dp$    \\
  $K^{*\pm} K^{\mp}$                  & $A+D+\Dp$     \\
  $\kstz\kazb$, $\kstzb \kaz$         & $A-2D+\Dp$    \\
  $\omega \eta$                       & $A-D+2\Dp$    \\
  $\omega \piz$                       & $~~\sqrt{3}D~~~$     \\
  $\roz \eta$                         & $~~\sqrt{3}D~~~$    \\
\hline \hline
\end{tabular}
\end{table}

\begin{table}[hbt]
\caption{\label{ppmsnform}  Amplitude parametrization form for
decays of the $\psp$ or $\jpsi$ into $P P$ final states. General expressions in terms of the charge-breaking one ($D=2g_e$) and the mass-breaking one ($\Dp=-2 g_m$).
The table can also be used for more similar decays by appropriate change in labeling, refer to appendix for more details.
} \center
\begin{tabular}{lc}\hline \hline
  Final state    & Amplitude parametrization form  \\ \hline
  $\pip\pim$     & $D$    \\
  $\kap\kam$     & $D+\Dp$     \\
  $\kaz\kazb$    & $~~~~~~\Dp$       \\
\hline \hline
\end{tabular}
\end{table}

In the light of Eq.~\eref{effhmtvpmsn}, the first term $\gz \cdot [V P]_0$ represents the $SU(3)$-conserved effect while in Eq.~\eref{effhmtppmsn} only the $SU(3)$-breaking terms exist. Therefore, for VP-like final states, the decay branching fractions are generally greater than those of PP-like final states.

In above analysis, it is prominent that in Eqs.~\eref{oktvtmsn} and \eref{oktpsmsn}, the mesons $\omega$ and $\eta$ are treated as pure octet components, but the real or observable $\omega$ and $\eta$ are actually the mixing of pure octet and singlet components with a certain mixing angle. According to PDG~\cite{pdg2020},
\beq
\left.\begin{array}{rcl}
\phi   &=&\omega^8~\cos \theta_V -\omega^1~\sin \theta_V~,   \\
\omega &=&\omega^8~\sin \theta_V +\omega^1~\cos \theta_V~,
\end{array}\right.
\label{mixvtmsn}
\eeq
or its reverse
\beq
\left.\begin{array}{rcl}
\omega^8  &=&\cos \theta_V ~\phi + \sin \theta_V~\omega~,   \\
\omega^1 &=&-\sin \theta_V ~\phi + \cos \theta_V~\omega~,
\end{array}\right.
\label{mixvtmsnrv}
\eeq
and
\beq
\left.\begin{array}{rcl}
\eta   &=&\eta^8~\cos \theta_P -\eta^1~\sin \theta_P~,   \\
\etap  &=&\eta^8~\sin \theta_P +\eta^1~\cos \theta_P~,
\end{array}\right.
\label{mixpsmsn}
\eeq
or its reverse
\beq
\left.\begin{array}{rcl}
\eta^8  &=&\cos \theta_P ~\eta + \sin \theta_P~\etap~,   \\
\eta^1 &=&-\sin \theta_P ~\eta + \cos \theta_P~\etap~.
\end{array}\right.
\label{mixpsmsnrv}
\eeq
In above equations, the superscript $8$ indicates the pure octet component and the superscript $1$ the pure singlet component.

In order to obtain the actual effective Hamiltonian, both octet and singlet components are to be included. The concise and effective way is to introduce a nonet that merges a singlet with an octet~\cite{Haber}. Under such circumstances, the matrixes in Eqs.~\eref{oktvtmsn} and \eref{oktpsmsn} are recast as

\beq
{\mathbf V_N}=
\left(\begin{array}{ccc}
V^1_1   & \rop    & \kstp    \\
\rom    & V^2_2   & \kstz    \\
\kstm   & \kstzb  &  V^3_3
\end{array}\right)~~,
\label{notvtmsn}
\eeq
with
\beq
\left.\begin{array}{rcl}
V^1_1&=&\roz/\sqrt{2}+\omega^8/\sqrt{6} +\omega^1/\sqrt{3}~,    \\
V^2_2&=& -\roz/\sqrt{2}+\omega^8/\sqrt{6} +\omega^1/\sqrt{3}~,    \\
V^3_3&=& -2 \omega^8/\sqrt{6}+\omega^1/\sqrt{3}~;
\end{array}\right.
\eeq
and
\beq
{\mathbf P_N}=
\left(\begin{array}{ccc}
P^1_1    & \pip         & \kap    \\
\pim     & P^2_2  & \kaz    \\
\kam     & \kazb  & P^3_3
\end{array}\right)~~,
\label{notpsmsn}
\eeq
with
\beq
\left.\begin{array}{rcl}
P^1_1&=&\piz/\sqrt{2}+\eta^8/\sqrt{6} +\eta^1/\sqrt{3} ~,   \\
P^2_2&=& -\piz/\sqrt{2}+\eta^8/\sqrt{6} +\eta^1/\sqrt{3}~, \\
P^3_3&=& -2 \eta^8/\sqrt{6} +\eta^1/\sqrt{3}~.
\end{array}\right.
\eeq
With these new nonets ${\mathbf V_N}$ and ${\mathbf P_N}$, the effective Hamiltonian in Eq.~\eref{effhmtvpmsn} can be used formally to acquire the
corresponding parametrization for ${\mathbf V} {\mathbf P}$ final states, with both octet components, say $\omega^8$ and $\eta^8$, and singlet components, say $\omega^1$ and $\eta^1$. Then with Eqs.~\eref{mixvtmsnrv} and \eref{mixpsmsnrv}, the pure octet and singlet components $\omega^8$, $\omega^1$, $\eta^8$, and $\eta^1$
will be replaced by the actual particles $\omega$, $\phi$, $\eta$, and $\etap$.
The final parametrization results are summarized in Table~\ref{vpmsnfmnew}.

\begin{table}[hbt]
\caption{\label{vpmsnfmnew} Amplitude parametrization form for
decays of the $\psp$ or $\jpsi$ into $V~P$ final states. General expressions in terms of singlet $A$ (by definition $A=g_0$), as well as charge-breaking ($D=g_e/\sqrt{3}$) and mass-breaking terms ($\Dp=2g_m/\sqrt{3}$). The shorthand symbols are defined as $s_{\alpha}=\sin \theta_{\alpha}$,  $c_{\alpha}=\cos \theta_{\alpha}$, $s^{\pm}_{\alpha\beta}=\sin ( \theta_{\alpha}\pm \theta_{\beta})$,$c^{\pm}_{\alpha\beta}=\cos ( \theta_{\alpha}\pm \theta_{\beta})$,$s_{\gamma \alpha\beta}=\sin ( \theta_{\gamma} +\theta_{\alpha}+\theta_{\beta})$, $c_{\gamma \alpha\beta}=\cos ( \theta_{\gamma} +\theta_{\alpha}+\theta_{\beta})$, $s_{\gamma}= \sin \theta_{\gamma}\equiv\sqrt{1/3}$, and $c_{\gamma}= \cos \theta_{\gamma}\equiv\sqrt{2/3}$. It should be noted that the definitions of $D$ and $\Dp$ herein are different from those in Table~\ref{vpmsnform}.} \center
\begin{tabular}{lccc}\hline \hline
  States    &  $A$        &   $D$        &  $\Dp$        \\ \hline
$\roz \piz$ &   1         &$ 1/\sqrt{3}$ &$ -1/\sqrt{3} $\\
$\rop \pim$ &   1         &$ 1/\sqrt{3}$ &$ -1/\sqrt{3} $\\
$\rop \pim$ &   1         &$ 1/\sqrt{3}$ &$ -1/\sqrt{3} $\\
$\kstp\kam$ &   1         &$ 1/\sqrt{3}$ &$ 1/2\sqrt{3} $\\
$\kstm\kap$ &   1         &$ 1/\sqrt{3}$ &$ 1/2\sqrt{3} $\\
$\kstz\kazb$&   1         &$-2/\sqrt{3}$ &$ 1/2\sqrt{3} $\\
$\kstzb\kaz$&   1         &$-2/\sqrt{3}$ &$ 1/2\sqrt{3} $\\
$\phi\eta$  &$c^{-}_{VP}$ &$-s_{\gamma VP}-s_{\gamma} s_V s_P$
                               &$s_{\gamma VP}+s_{\gamma} s_V s_P$  \\
$\phi\etap$ &$-s^{-}_{VP}$&$c_{\gamma VP}+s_{\gamma} s_V c_P$
                               &$-c_{\gamma VP}-s_{\gamma} s_V c_P$  \\
$\omega\eta$&$s^{-}_{VP}$ &$c_{\gamma VP}+s_{\gamma} c_V s_P$
                               &$-c_{\gamma VP}-s_{\gamma} c_V c_P$  \\
$\omega\etap$&$c^{-}_{VP}$&$s_{\gamma VP}-s_{\gamma} c_V s_P$
                               &$-s_{\gamma VP}+s_{\gamma} c_V c_P$  \\
$\roz\eta$  &   0    &$\sqrt{3}\cdot s^{-}_{\gamma P}$ &   0          \\
$\roz\etap$ &   0    &$\sqrt{3}\cdot c^{-}_{\gamma P}$ &   0          \\
$\phi\piz$  &   0    &$\sqrt{3}\cdot s^{-}_{\gamma V}$ &   0          \\
$\omega\piz$&   0    &$\sqrt{3}\cdot c^{-}_{\gamma V}$ &   0          \\
\hline \hline
\end{tabular}
\end{table}

Besides the nonet approach, the singlet component can be treated separately, and the final results are essentially the same. The details are degenerated into appendix.

\section{Experimental section}\label{xct_expmsm}
Electron-positron collider experiment has its special character. When analyzing the data taken in $\EE$ collider, the important experimental effects such as the initial state radiative (ISR) correction and the effect due to energy spread of accelerator must be dealt with carefully.

\subsection{Born section}

For $\EE$ colliding experiments, there is the inevitable continuum amplitude~\cite{wangp03hepnp}
$$ 
\EE \rightarrow \gamma^* \rightarrow hadrons
$$ 
which may produce the same final state as the resonance does.
In $\EE\rightarrow {\mathbf V} {\mathbf P}$ at $\jpsi$ or $\psp$ resonance, the Born order
cross section for final state $f$ is~\cite{rudaz,wymcgam,Wang:2005sk,wymphase,wymogpiapp}
\beq
\sigma_{Born} =
\frac{4\pi\alpha^2}{s^{3/2}}|A_f(s)|^2{\cal P}_f(s)~~,
\label{Born}
\eeq
where ${\cal P}_{f}(s) = q_f^3/3$, with $q_f$ being the momentum of either the
${\mathbf V}$ or the $ {\mathbf P}$ particle, viz.
\beq
q_f = \frac{[(s-(m_1+m_2)^2)(s-(m_1-m_2)^2)]^{1/2}}{2\sqrt{s}}~,
\label{qfexprn}
\eeq
where $\sqrt{s}$ is the center of mass energy, $m_1$ and $m_2$ are the masses of two ${\mathbf V} {\mathbf P}$ final state mesons. The total amplitude reads
\beq
A_f(s) = \ag(s)+\aga(s)+\ac(s) ~,
\label{amptot}
\eeq
which consists of three kinds of amplitudes corresponding to (a) the strong interaction [$\ag(s)$] presumably through three-gluon annihilation, (b) the electromagnetic interaction [$\aga(s)$]
through the annihilation of $c\overline{c}$ pair into a virtual photon, and (c) the electromagnetic interaction [$\ac(s)$] due to one-photon continuum process. For ${\mathbf V} {\mathbf P}$ final state, the amplitudes have the forms~:
\beq
\ac(s)=Y_f \cdot {\cal F}(s)~,
\label{ampac}
\eeq
\beq
\aga(s)=Y_f \cdot B(s) \cdot {\cal F}(s)~,
\label{ampap}
\eeq
\beq
\ag(s)=X_f \cdot B(s) \cdot {\cal F}(s)~,
\label{ampag}
\eeq
with the definition
\beq
B(s) \equiv \frac{3\sqrt{s}\Gamma_{ee}/\alpha}{s-M^2+iM\Gamma_t}~~,
\label{defbsfcn}
\eeq
where $\alpha$ is the QED fine structure constant; $M$ and $\Gamma_t$ are the mass and the total width of the $\psp$ or $\jpsi$; $\Gamma_{ee}$ is the partial width to $\EE$. ${\cal F}(s)$ is the form factor and take the form $1/s$ for ${\mathbf V} {\mathbf P}$ final state. $X_f$ and $Y_f$ are the functions of the amplitude parameters of final state $f$, that is $A,D$, $\Dp$, $s_P$, and $s_V$, \`{a} la Table~\ref{vpmsnfmnew}, viz.
\beq
Y=Y(D,s_P, s_V)~,
\label{defy}
\eeq
\beq
X=X(A,\Dp,s_P, s_V) e^{i\phi}~.
\label{defx}
\eeq
By virtue of Eq.~\eref{mixpsmsn} and Eq.~\eref{mixvtmsn}, $s_P=\sin \theta_P$ and $s_V=\sin \theta_V$, where $\theta_P$ and $\theta_V$ are respectively the mixing angle of pseudoscalar meson between $\eta$ and $\etap$, and that of vector meson between $\phi$ and $\omega$. The concrete form of $X$ or $Y$ depends on the decay mode, as examples, for $\rho \pi$ final state, $X_{\rho \pi}=A-\Dp/\sqrt{3}$ and $Y_{\rho \pi}=D/\sqrt{3}$ while for $\omega\piz$ final state, $X_{\omega\piz}=0$ and $Y_{\omega\piz}=D\cdot \sqrt{3} \cdot c^{-}_{\gamma V}$, according to the parametrization form in Table~\ref{vpmsnfmnew}. In principle, the parameters $A,D$, and $\Dp$ could be complex arguments, each with a magnitude together with a phase. Conventionally, it is assumed that there is not relative phases among the strong-originated amplitudes $A$, $\Dp$, and the sole phase [denoted by $\phi$ in Eq.~\eref{defx} ] is between the strong and electromagnetic interactions, that is between $X$ and $Y$, as indicated in Eqs.~\eref{defx} and \eref{defy}, where $A$, $D$, and $\Dp$ are actually treated as real numbers.

\begin{table*}[bth]
\caption{\label{tab_expcdn}Breakdown of experiment conditions correspond to different detectors and accelerators. The energy spread is the effective one, according to which the calculated maximum cross section satisfies the relation $N_{tot} =\sigma_{max} \cdot {\cal L}$. The number with star ($\ast$) is the equivalent luminosity calculated by the relation ${\cal L}=N_{tot}/\sigma_{max}$. }
\begin{ruledtabular}
\begin{tabular}{llllllll}
         &         &Center of Mass Energy &Data Taking & Maximum   & Total & Integral   & \\
Detector & Accelerator &  Spread    &Position\footnote{
\begin{minipage}{13cm}\mbox{}The data taking position is the energy which
yield the maximum inclusive hadronic cross section. \end{minipage} }
                                                 & section   & event & luminosity &References \\
         &             & (MeV)      & (GeV)      &  (nb)     &($\times 10^6$)
                                                                     & (pb$^{-1}$) & \\ \hline
BES II  & BEPC    &   1.23  & 3.68623   & 712.9    &$14.0\pm 0.6$  & 19.72      & \cite{tnpsp2004}\\
        &         &   1.23  & 3.65      & $\cdots$ &$\cdots$ & 6.42       & \cite{lum2004}\\
BESIII  & BECPII  &   1.343 & 3.68624   & 662.16   &$107.0\pm 0.8$ & 161.63     & \cite{tnpsp2018}\\
        &         &   1.343 & 3.65      & $\cdots$ &$\cdots$ & 43.88      & \cite{tnpsp2018} \\
        &         &   1.318 & 3.68624   & 672.74   &$341.1\pm 2.1$ & 506.92     & \cite{tnpsp2018} \\
        &         &   1.324 & 3.68624   & 670.17   &$448.1\pm 2.9$ & 668.55     & \cite{tnpsp2018}\\
        &         &   1.324 & 3.65      & $\cdots$ &$\cdots$       & 48.8      & \cite{besretp2017} \\
CLEO-c  & CESR    &   1.68  & 3.68627   & 557.23   & 3.08    & 5.63       & \cite{cleovp2005} \\
        &         &   1.68  & 3.67      & $\cdots$ & $\cdots$    & 20.46      & \\
BESIII   & BECPII & 1.131   & 3.097014  & 2808.63  &$223.7\pm 1.4$ & 79.63  & \cite{tnjps2012}\\
            &     & 1.131   & 3.08      & $\cdots$ & $\cdots$      & 0.282  & \cite{bes3pi2012} \\
            &     & 0.898   & 3.096990  & 3447.87  &$1086.9\pm 6.0$& 315.02 & \cite{tnjps2017} \\
            &     & 0.937   & 3.096993  & 3320.35  &$1310.6\pm 7.0$& 394.65  & \cite{tnjps2017} \\
            &     & 0.937   & 3.08      &$\cdots$  &$\cdots$       & 153.8   & \cite{besretp2017} \\
BES II   & BEPC   & 0.85    & 3.09700   & 3631.8   &$57.7\pm 2.72$ & 15.89$\ast$ & \cite{tnjps2003}\\
            &     & 0.85    & 3.07      &$\cdots$  &$\cdots$       & 2.3473  & \cite{besjopaet06} \\
 DM  II  & DCI    & 1.98    & 3.097114  & 1702.0    &$8.6\pm 1.3$  & 5.053$\ast$ & \cite{JJousset90jvp}\\
\end{tabular}
\end{ruledtabular}
\end{table*}

\subsection{Observed section}

In $\EE$ collision, the Born order cross section is
modified by the initial state radiation in the way~\cite{rad.1}
\begin{equation}
\sigma_{r.c.} (s)=\int \limits_{0}^{x_m} dx
F(x,s) \frac{\sigma_{Born}(s(1-x))}{|1-\Pi (s(1-x))|^2},
\label{eq_isr}
\end{equation}
where $x_m=1-s'/s$. $F(x,s)$ is the radiative function been calculated to an accuracy of
0.1\%~\cite{rad.1,rad.2,rad.3}, and $\Pi(s)$ is the vacuum polarization factor. In the
upper limit of the integration, $\sqrt{s'}$ is the experimentally required minimum
invariant mass of the final particles. If $x_m=1$, it corresponds to no requirement for invariant mass; if $x_m=0.2$, it corresponds to invariant mass cut of 3.3~GeV for $\psp$ resonance.
The concrete value of $x_m$ should be determined by the cut of invariant mass, which is adopted in actual event selection.

By convention, $\Gamma_{ee}$ has the QED vacuum polarization in its definition~\cite{Tsai,Luth}. Here it is natural to extend this convention to the partial widths of other pure electromagnetic decays, that is
\beq
\Gamma_{f} = \frac{\tilde{\Gamma}_{ee} q^3_{f}}{M}
\left|{\cal F} (M^2) \right|^2~,
\label{eq_defgf}
\eeq
where
$$ \tilde{\Gamma}_{ee} \equiv
\frac{\Gamma_{ee}}{|1-\Pi (M^2)|^2}~$$
with the vacuum polarization effect included.

The $\EE$ collider has a finite energy resolution which is much wider than the intrinsic width of narrow
resonances such as $\psp$ and $\jpsi$~\cite{besscan95,besscan02}. Such an energy resolution is usually a Gaussian distribution~\cite{awChao}:
$$
G(W,W^{\prime})=\frac{1}{\sqrt{2 \pi} \Delta}
             e^{ -\frac{(W-W^{\prime})^2}{2 {\Delta}^2} },
$$
where $W=\sqrt{s}$ and $\Delta$, a function of energy, is the standard deviation of Gaussian distribution. The experimentally observed cross section is the radiative corrected cross section folded with the energy resolution function
\begin{equation}
\sigma_{obs} (W)=\int \limits_{0}^{\infty}
        dW^{\prime} \sigma_{r.c.} (W^{\prime}) G(W^{\prime},W)~.
\label{eq_engsprd}
\end{equation}

In fact, as pointed out in Ref.~\cite{wymcgam}, the radiative correction and the energy spread of the collider are two important factors, both of which reduce the height of the resonance and shift the position of the maximum cross section. Although the ISR are the same for all $\EE$ experiments, the energy spread is quite different for different accelerators, even different for the same accelerator at different running periods. As an example, for the CLEO data used in this paper, the energy spread varies due to different accelerator lattices~\cite{YELLOWBOOK}: one (for CLEO III detector) with a single wiggler
magnet and a center-of-mass energy spread $\Delta$=1.5~MeV, the other (for CLEOc detector) with the first half of its full complement (12) of wiggler magnets and $\Delta$=2.3~MeV. The two $\Delta$'s lead to two maximum total cross sections 602 nb and 416 nb, respectively. All these subtle effects must be taken into account in data analysis. In the following analysis all data are assumed to be taken at the energy point which yields the maximum inclusive hadron cross sections in stead of the nominal resonance mass~\cite{wymcgam,wymhepnp}. Besides the factors considered above, the resonance parameters can also affect the evaluation results. Since the present central values of resonance parameters can be obviously distinct from those of some time before, the calculated maximum inclusive hadron cross sections will consequently different. In order to ensure the relation $N_{tot} =\sigma_{max} \cdot {\cal L}$, some adjustments are mandatory. The principle is as follows: if the luminosity is available, the energy spread will be tuned to give consistent maximum cross section; otherwise, the effective luminosity is evaluated by the relation ${\cal L}=N_{tot}/\sigma_{max}$ according to the corresponding accelerator parameters. All relevant experimental details are summarized in Table~\ref{tab_expcdn}, which are crucial for the following data analysis. At last, the resonance parameters adopted in this paper for $\jpsi$ and $\psp$ are respectively~\cite{pdg2020}
\beq
\begin{array}{rcl}
   M_R &=&3096.900\pm 0.006 ~~\mbox{MeV }, \\
   \Gamma_t &=&92.9\pm 2.8~~\mbox{keV }, \\
   \Gamma_{ee}&=& 5.53\pm 0.10~~\mbox{keV };
\end{array}
\label{jpsirnsprt}
\eeq
and
\beq
\begin{array}{rcl}
   M_R &=&3686.10\pm 0.06 ~~\mbox{MeV }, \\
   \Gamma_t &=&294\pm 8~~\mbox{keV }, \\
   \Gamma_{ee}&=& 2.33\pm 0.04~~\mbox{keV }.
\end{array}
\label{psiprnsprt}
\eeq

\begin{table*}[bth]
\caption{\label{tab_psptovp}Experimental data of $\psp$ decaying to $V P$ final states.
For branching ratios, the first uncertainties are statistical, and the second systematic. The peak position is assumed at $\sqrt{s}=3.686$ GeV, while the continuum position is assumed at $\sqrt{s}=3.65$ GeV for BES and $\sqrt{s}=3.67$ GeV for CLEO. The efficiency indicates the selection efficiency for resonance events. The number in parenthesis is not quoted in the original literature but is evaluated according to the information given therein.}
\begin{ruledtabular}
\begin{tabular}{llllll}
  Mode  & $N^{obs}$ & $N^{obs}$  & Efficiency   & Branching Ratio        & Detector     \\
        &  (peak) & (continuum)  & (\%)         &     ($\times 10^{-5}$) &     \\ \hline
$\pip\pim\piz$
        &$7771\pm 88$&$220.6\pm14.8$ &$30.5$
                                    &$2.14\pm 0.03^{+0.08}_{-0.07} $& BESIII~\cite{bes3pi2012} \\
        &$216.7$    &$85    $ &$33.5$ &$18.8^{+1.6}_{-1.5} \pm 2.8 $& CLEO~\cite{cleovp2005} \\
        &$260\pm 19$&$10.0\pm4.2$ &$9.02$ &$18.1 \pm 1.8 \pm 1.9   $& BESII~\cite{besrp2005} \\
$\rpi$  &$34.4 $    &$47    $ &$28.8$ &$2.4^{+0.8}_{-0.7} \pm 0.2  $& CLEO~\cite{cleovp2005} \\
        &$64.12\pm 6.44$ &$\cdots$ &$9.02$ &$5.1\pm 0.7 \pm 1.1    $& BESII~\cite{besrp2005} \\
$\kstp\kam+c.c.$
        &$7.7  $    &$4     $ &$16.7$ &$1.3^{+1.0}_{-0.7} \pm 0.3  $& CLEO~\cite{cleovp2005} \\
        &$9.6\pm 4.2$&$\cdots$&$7.3$  &$2.9^{+1.3}_{-1.7} \pm 0.4  $& BESII~\cite{beskstk2005} \\
        &$224\pm 21$ &$(54.7\pm 7.6)$
                              &$20.25$&$3.18\pm 0.30^{+0.26}_{-0.31}$& BESIII~\cite{besvps2012} \\
$\kstz\kazb+c.c$
        &$34.5 $    &$36    $ &$8.7 $ &$9.2^{+2.7}_{-2.2} \pm 0.9  $& CLEO~\cite{cleovp2005} \\
        &$65.6\pm 9.0$&$2.5^{+2.6}_{-1.8}$
                              &$9.7$  &$12.3^{+2.4}_{-2.6} \pm 1.7 $& BESII~\cite{beskstk2005} \\
$\phi\eta$
        &$6.6 $  &$3    $  &$9.4 $  &$2.0^{+1.5}_{-1.1} \pm 0.4    $& CLEO~\cite{cleovp2005} \\
        &$16.7\pm 5.6$&$\cdots$&$18.9$  &$3.3\pm 1.1 \pm 0.5       $& BESII~\cite{besfw2004} \\
        &$216\pm 16$  &$(7.0\pm 2.5)$
                           &$33.53$ &$3.14\pm 0.23 \pm 0.23        $& BESIII~\cite{besvps2012} \\
$\phi\etap$
        &$8.4\pm 3.7$&$\cdots$&$8.4$   &$3.1\pm 1.4 \pm0.7         $& BESII~\cite{besfw2004} \\
        &$201\pm 15 $&$221\pm 15$\footnote{
\begin{minipage}{7.5cm}\mbox{}The number of event is obtained at $\sqrt{s}=3.773$ GeV. \end{minipage} }
                           &$26.8$ &$1.51\pm 0.16 \pm0.12 $& BESIII~\cite{besfet2019} \\
$\omega\eta$&$<0$   &$3   $ &$10.2 $   &$<1.1                      $& CLEO~\cite{cleovp2005} \\
        &$<9.7$     &$\cdots$&$6.3$ &$<3.1                         $& BESII~\cite{besfw2004} \\
$\omega\etap$
        &$4.2^{+3.2}_{-2.7}$&$\cdots$&$2.3$ &$3.2^{+2.4}_{-2.0} \pm 0.7 $& BESII~\cite{besfw2004} \\
$\roz\eta$ &$28.1  $&$38   $ &$19.3 $ &$3.0^{+1.1}_{-0.9} \pm 0.2  $& CLEO~\cite{cleovp2005} \\
        &$29.2^{+7.5}_{-6.8}$&$2.3^{+2.1}_{-1.4}$ &$(12.06)$
                                  &$1.87^{+0.68}_{-0.62} \pm 0.18  $& BESII~\cite{besret2004} \\
$\roz\etap$
        &$5.4^{+3.3}_{-2.2}$&$<4.4$ &$(4.92)$
                                  &$1.87^{+1.64}_{-1.11} \pm 0.33  $& BESII~\cite{besret2004} \\
&($211\pm 16$)\footnote{
\begin{minipage}{11.00cm}\mbox{}The solution of destructive interference between $\rho$ and nonresonant components.  \end{minipage} }
                &$5.06\pm2.01$  &$18.7 $ &$1.02\pm 0.11 \pm 0.24 $  & BESIII~\cite{besretp2017}\\
&($148\pm 18 $)\footnote{
\begin{minipage}{11.10cm}\mbox{}The solution of constructive interference between $\rho$ and nonresonant components.  \end{minipage} }
                &$5.06\pm2.01$  &$18.7 $ &$0.569\pm 0.128\pm 0.236$ & BESIII~\cite{besretp2017}\\
$\phi\piz$  &$<0$   &$3   $ &$15.8 $   &$<0.7                       $& CLEO~\cite{cleovp2005} \\
            &$<4.4$ &$\cdots$&$16.1 $   &$<0.4                       $& BESII~\cite{besfw2004} \\
            &$<6$   &$\cdots$&$35.63 $  &$<0.04                      $& BESIII~\cite{besvps2012} \\
$\omega\piz$  &$29.1  $&$55 $  &$19.1 $   &$2.5^{+1.2}_{-1.0} \pm 0.2  $& CLEO~\cite{cleovp2005} \\
        &$31.2^{+7.7}_{-6.9}$&$7.3^{+3.3}_{-2.7}$ &$(5.45)$
                                  &$1.78^{+0.67}_{-0.62} \pm 0.28  $& BESII~\cite{besret2004} \\
\end{tabular}
\end{ruledtabular}
\end{table*}

\section{Data analysis}\label{xct_fsfit}

The Standard Model mainly consists of two parts. One is Salam-Weinberg model that depicts the electroweak interaction, which can usually accommodate accurate enough evaluation for certain process. Another part is QCD, the validity of which at nonperturbative regime needs more experimental guidance. The production and decay of charmonium states benefit such a study.

As one of important and interesting steps, it is a good start point to study the relative phase between the strong and electromagnetic (EM) interaction amplitudes, which provides us a new viewpoint to explore the quarkonium decay dynamics and profound our understanding on QCD. Studies have been carried out for many $\jpsi$ and $\psp$ two-body mesonic decay modes with various spin-parities~: $1^-0^-$~\cite{wymphase,JJousset90jvp,mk3jvp85,mk3jvp88}, $0^-0^-$~\cite{a00,LopezCastro:1994xw,wymppdk,bes2klks04,a11}, and $1^-1^-$~\cite{a11}. These analyses reveal that there exists a relative orthogonal phase between the EM and strong decay amplitudes. There is also a conjecture to claim that such an orthogonal phase is universal for all quarkonia decays~\cite{Wang:2003zx,gerard}.

The systematical parametrization scheme of charmonium decay modes facilitates the study of the relative phase. In Ref.~\cite{moxh2024}, the phase is measured for various charmonium decay modes, including $\psp$ and/or $\jpsi$ decay to octet baryon pair, decuplet baryon pair, decuplet-octet baryon final state, and pseudoscalar-pseudoscalar meson final state. In this section, the study is devoted to vector-pseudoscalar (${\mathbf V} {\mathbf P}$) meson final state according to the parametrization of Table~\ref{vpmsnfmnew}.

Since our analysis involves the experimental details as indicated by the description in preceding section, some measurements are not suitable in the following study due to the lack of necessary information of detectors and/or accelerators. In addition, at different energy point, the status parameters of accelerators are also distinctive, so the studies of phase angle for $\psp$ and $\jpsi$ decay are performed separately for the sake of clarity.

\begin{table}[bth]
\caption{\label{tab_vpdata}Data of $\psp \to VP$ decays from CLEO and BES experiments. The error is merely the statistical and the efficiency is the effective one as depicted in the text.}
\begin{ruledtabular}
\begin{tabular}{lllll}
  Mode  & Energy  &   $N^{obs}$      &Efficiency & Detector     \\
        &  (GeV)  &                  &  (\%)     & \\ \hline
$\rpi$  &$3.686 $ &$54.24\pm 11.53$  &$33.5$  & CLEO~\cite{cleovp2005} \\
        &$3.686 $ &$64.12\pm 6.44$   &$9.02$  & BESII~\cite{besrp2005} \\
$\kstp\kam+c.c.$
        &$3.686 $ &$7.7\pm 5.9  $    &$16.7$  & CLEO~\cite{cleovp2005} \\
        &$3.67  $ &$4 \pm 3.1  $     &$16.7$  &   \\
        &$3.686 $ &$9.6\pm 4.2 $     &$2.34$   & BESII~\cite{beskstk2005} \\
        &$3.686 $ &$224\pm 21 $      &$6.65$   & BESIII~\cite{besvps2012} \\
$\kstz\kazb+c.c$
        &$3.686 $ &$34.5\pm 10.1 $   &$8.7 $  & CLEO~\cite{cleovp2005}  \\
        &$3.67  $ &$36\pm 10.6   $   &$8.7 $  &  \\
        &$3.686 $ &$65.6\pm 9.0 $    &$3.11$   & BESII~\cite{beskstk2005} \\
        &$3.65  $ &$2.5 \pm 2.6 $    &$3.11$  &   \\
$\phi\eta$
        &$3.686 $ &$6.6\pm 5.0 $     &$9.4 $     & CLEO~\cite{cleovp2005} \\
        &$3.67  $ &$3\pm 2.3   $     &$9.4 $     &  \\
        &$3.686 $ &$16.7\pm 5.6$     &$3.67$     & BESII~\cite{besfw2004} \\
        &$3.686 $ &$216\pm 16$       &$6.50$    & BESIII~\cite{besvps2012} \\
$\phi\etap$
        &$3.686 $ &$8.4\pm 3.7$      &$1.92$      & BESII~\cite{besfw2004} \\
        &$3.686 $ &$201\pm 15 $      &$2.21$     & BESIII~\cite{besfet2019} \\
$\omega\eta$
        &$3.686 $ &$4.2\pm 3.2$      &$0.95 $     & BESII~\cite{besfw2004} \\
$\roz\eta$
        &$3.686 $ &$28.1\pm 10.3  $  &$19.3 $     & CLEO~\cite{cleovp2005}  \\
        &$3.67  $ &$38 \pm 13.9  $   &$19.3 $     &  \\
        &$3.686 $ &$29.2\pm 7.5  $   &$4.75$   & BESII~\cite{besret2004} \\
        &$3.65  $ &$2.3\pm 2.1   $   &$4.29$   &  \\
$\roz\etap$
        &$3.686 $ &$5.4\pm  3.3$     &$0.86$   & BESII~\cite{besret2004} \\
&$3.686 $&($211\pm 16$)\footnote{\begin{minipage}{3.50cm}\mbox{}The destructive solution.\end{minipage} }
                                     &$3.116 $ & BESIII~\cite{besretp2017}\\
&$3.686 $&($148\pm 18 $)\footnote{\begin{minipage}{3.60cm}\mbox{}The constructive solution.   \end{minipage} }                     &$3.116 $ & BESIII~\cite{besretp2017}\\
        &$3.65  $ &$5.06\pm2.01$     &$3.116 $ & BESIII~\cite{besretp2017}\\

$\omega\piz$
        &$3.686 $ &$29.1\pm 14.0  $  &$19.1 $     & CLEO~\cite{cleovp2005} \\
        &$3.67  $ &$55 \pm 26.4 $    &$19.1 $     &  \\
        &$3.686 $ &$31.2\pm 7.7   $  &$4.87 $  & BESII~\cite{besret2004}  \\
        &$3.65  $ &$7.3\pm  3.3  $   &$4.55 $  &  \\
\end{tabular}
\end{ruledtabular}
\end{table}

\subsection{$\psp \to VP$ decay }
There are lots of measurements concerned with $\psp$ decaying to ${\mathbf V} {\mathbf P}$ final states. 
However, the results of Ref.~\cite{Franklin83} are obtained forty years ago, and moreover only the upper limits of $K^{*\pm}K^{\mp}$ and $\RP$ final states are given based on 1 million $\psp$ events. The results from Ref.~\cite{wBartel76} are merely the upper limits for $\RP$ and $\gamma \eta^{\prime}$ final states based on 0.2 million $\psp$ events. Therefore, these kinds of measurements are not adopted in our study.  The results of experimental measurements obtained in this century are collected in Table~\ref{tab_psptovp}, which are mainly due to CLEO and BES collaborations.

First, let's focus on $\rpi$ final state. By virtue of Table~\ref{tab_psptovp}, the branching fraction of $\psp \to \pip\pim\piz$ agrees very well for both CLEO and BES experiments, while that of $\psp \to \rpi$ are rather different. As pointed out in Ref.~\cite{besrp2005}, the partial wave analysis indicates that in all $\pip\pim\piz$ events, only $28\%$ are due to $\rpi$ final state. It seems that the result from BES takes more information into account. If we adopt the same proportion of $\rpi$ for CLEO data, the branching fraction $\psp \to \rpi$ is around $5.3 \times 10^{-5}$, which is fairly consist with BES result. Therefore, the modified data from CLEO are adopted in the following analysis, the details are presented in Table~\ref{tab_vpdata}.

Second, in Ref.~\cite{besretp2017} based on 448 million $\psp$ events, $\pip\pim\etap$ final state is studied resorting to partial wave analysis technique. The interference between $\rho$ and nonresonant components is observed. The constructive and destructive interferences lead to two possible solutions of branching fraction, that is $(5.69\pm 1.28\pm 2.36) \times 10^{-6}$ and $(1.02\pm 0.11 \pm 0.24)\times 10^{-5}$, respectively. In the following fit, two results will be dealt with separately.

Third, for many ${\mathbf V} {\mathbf P}$ decay modes, there are the intermediate states. Take $\phi \eta$ mode as an example, $\phi \to \kk$ and $\eta \to \GG$, in calculation of branching ratio, the intermediate decay branching ratios must be taken into consideration. Such kinds of effects could also be included in Monte Carlo simulation. Therefore, it must be careful to figure out how such kinds of effects are taken into account. Moreover, there are some efficiency correction factors due to detector or Monte Carlo simulation, which should also be considered. In a word, the efficiency in the following analysis is the one that includes all kinds of necessary effects, and is termed as the effective efficiency.

Fourth, as far as the aforementioned principle is concerned, the energy spread will be tuned to give the maximum cross section that can satisfy the relation $N_{tot} =\sigma_{max} \cdot {\cal L}$. CLEO data~\cite{cleovp2005} are composed of two sets, one with luminosity 2.74 pb$^{-1}$ and the other 2.89 pb$^{-1}$, which are taken with energy spreads 1.5~MeV and 2.3~MeV, respectively. In the following analysis, the data is treated as one set with total luminosity 5.63 pb$^{-1}$ corresponding to the effective energy spread 1.68~MeV as displayed in Table~\ref{tab_expcdn}. It it worthy of noticing that unlike branching fraction evaluation, the contribution due to QED continuum should not be subtracted from the observed number of events, since the QED contribution is included in the calculation of observed cross section. At last, since the error of number of events are needed in analysis, for CLEO data, the maximum relative statistical error of branching fraction is used to evaluate the corresponding error of the number of events.

Chi-square method is adopted to fit the experimental data. The estimator is constructed as
\beq
\chi^2= \sum\limits_i
\frac{[N_i - n_i(\vec{\eta})]^2}{(\delta N_i)^2}~,
\label{chisqbb}
\eeq
where $N$ with the corresponding error ($\delta N$) denotes the experimentally measured number of events while $n$ the theoretically calculated number of events~:
\beq
n={\cal L} \cdot \sigma_{obs} \cdot \epsilon~,
\label{eq_defsig}
\eeq
where ${\cal L}$ is the integrated luminosity, $\epsilon$ is the effective efficiency, and $\sigma_{obs}$ the observed cross section calculated according to formula~\eref{eq_engsprd}, which contains the 
parameters to be fit, such as $A$, $D$, $\Dp$, $s_P$, $s_V$,
and the phase angle $\phi$. All these parameters are denoted by the parameter vector $\vec{\eta}$ in Eq.~\eref{chisqbb}. The concrete form is determined by the parametrization form in Table~\ref{vpmsnfmnew}.
All observed numbers of events together with the corresponding efficiencies displayed in Table~\ref{tab_vpdata} are employed as input information. The data can be grouped into four sets: two from BESIII, one with total luminosity 668.55 pb$^{-1}$, the other with luminosity 161.63 pb$^{-1}$; one from CLEO, with total luminosity 5.63 pb$^{-1}$ and one from BESII with luminosity 19.72 pb$^{-1}$. There might be some systematic difference among those datasets, so normalization factors are introduced to take into account these systermatic effects. However, only three relative (relative to the greatest data set of BESIII) factors of luminosity are introduced with the belief that the relative relations of measurements of each experiment group is more reliable than the corresponding absolute values. The fit values of three factors $f_{cleo}$, $f_{bes2}$, and $f_{bes3a}$ indicate that there indeed exists certain obvious differences,
the inconsistencies of these experiments from the highest precision one range from 10\% to 70\%.

The fitting yields a $\chi^2$ of 18.91 with the number of degrees of freedom being 19 for the destructive inference case~:
\beq
\begin{array}{rcl}
  \phi    &=&-131.55^\circ \pm  13.05^\circ, \\
      A   &=&~0.577\pm 0.053~, \\
      D   &=&~0.334\pm 0.026~, \\
    \Dp   &=&-0.025\pm 0.078~,  \\
    s_P   &=&-0.277\pm 0.055~, \\
  \theta_P&=&-16.10^\circ~ , \\
    s_V   &=&~0.279\pm 0.195~, \\
  \theta_V&=&~16.19^\circ~, \\
 f_{cleo} &=&~0.937\pm 0.167~,  \\
 f_{bes2} &=&~1.277\pm 0.208~, \\
 f_{bes3a}&=&~1.361\pm 0.252~;
\end{array}
\label{ftvpa}
\eeq
and a $\chi^2$ of 15.82 with the number of degrees of freedom being 19 for the constructive inference case~:
\beq
\begin{array}{rcl}
  \phi    &=&-144.31^\circ \pm  20.93^\circ, \\
      A   &=&~0.545\pm 0.047~, \\
      D   &=&~0.300\pm 0.027~, \\
    \Dp   &=&-0.030\pm 0.068~,  \\
    s_P   &=&-0.307\pm 0.058~, \\
  \theta_P&=&-17.88^\circ~ , \\
    s_V   &=&~0.386\pm 0.221~, \\
  \theta_V&=&~22.73^\circ~, \\
 f_{cleo} &=&~1.141\pm 0.220~,  \\
 f_{bes2} &=&~1.516\pm 0.264~, \\
 f_{bes3a}&=&~1.703\pm 0.364~;
\end{array}
\label{ftvpb}
\eeq
where $\theta_\alpha = \arcsin s_\alpha~(\alpha=P,V)$.

The scan for parameter $\phi$ displays merely one minimum, which is a rather uncommon case for multiple solution theory. When fitting cross sections with several resonances or interfering background and resonances, one usually obtains multiple solutions of parameters with equal fitting quality. Such a phenomenon was firstly noticed experimentally~\cite{moxh2010prd,yuancz2010ijmpa}, then some studies are performed from a mathematical point of view~\cite{yuancz2011cpc,zhuk2011ijmpa,hanx2018cpc,baiyu2019prd}. Especially in Ref.~\cite{baiyu2019prd}, the source of multiple solutions for a combination of several resonances or interfering background and resonances is found by analyzing the mathematical structure of the Breit-Wigner function. It is proved that there are exactly $2^{n-1}$ fitting solutions with equal quality for $n$ amplitudes, and the multiplicity of the interfering background function and resonance amplitudes depends on zeros of the amplitudes in the complex plane. Our study involves the interference between strong and electromagnetic amplitudes, corresponding to $n=2$ case, therefore it is expected that there are two solutions for the phase angle $\phi$. For the present fit result, we are not sure if this is the special feature of $\psp \to {\mathbf V} {\mathbf P}$ decay, or the current data are not accurate enough to differentiate two solutions which are close to each other.

Besides the fit for all data in Table~\ref{tab_vpdata}, we also perform the fit for part of it. In the light of parametrization of Table~\ref{vpmsnfmnew}, it is obvious that $\roz\eta$, $\roz\etap$, $\phi\piz$, and $\omega\piz$ are only concerned with $D$, $s_P$, and $s_V$, therefore the fit of the data relevant to these final states can be used to determine the three parameters. Based on the information of Table~\ref{tab_vpdata} related to $\roz\eta$, $\roz\etap$, and $\omega\piz$ final states, the fitting yields a $\chi^2$ of 7.08 with the number of degrees of freedom being 8 for the destructive inference case~:
\beq
\begin{array}{rcl}
      D   &=&~0.369\pm 0.023~, \\
    s_P   &=&-0.335\pm 0.044~, \\
  \theta_P&=&-19.57^\circ~ , \\
    s_V   &=&~0.853\pm 0.136~, \\
  \theta_V&=&~58.54^\circ~;
\end{array}
\label{ftvpas}
\eeq
and a $\chi^2$ of 4.68 with the number of degrees of freedom being 8 for the constructive inference case~:
\beq
\begin{array}{rcl}
      D   &=&~0.351\pm 0.025~, \\
    s_P   &=&-0.405\pm 0.044~, \\
  \theta_P&=&-23.90^\circ~ , \\
    s_V   &=&~0.342\pm 0.400~, \\
  \theta_V&=&~20.02^\circ~.
\end{array}
\label{ftvpbs}
\eeq
It can be seen that the fit value of $\theta_P$ is marginally consist with the overall fit, which means that the value of $\theta_P$ is dominantly determined by $\roz\eta$ and $\roz\etap$ final states. On the contrary, since only the data of $\omega\piz$ final state is available, the fluctuation of fit value of $\theta_V$ is fairly prominent. Anyway, if the precise measurements of $\roz\eta$, $\roz\etap$, $\phi\piz$, and $\omega\piz$ final states can be obtained, the mixing angle of $\theta_P$ and $\theta_V$ is expected to be determined definitely.

\begin{table*}[bth]
\caption{\label{tab_jpstovp}Experimental data of $\jpsi$ decaying to $V P$ final states. For branching ratios, the first uncertainties are statistical, and the second systematic. The peak position is assumed at $\sqrt{s}=3.097$ GeV, while the continuum position is assumed at $\sqrt{s}=3.07$ or  $\sqrt{s}=3.08$ GeV for BES measurement. The effective efficiency is presented and the one with star ($\ast$) is evaluated by virtue of the observed number of events $N^{obs}$, the total number of resonance events, and the corresponding of branching fraction. The symbol ``$n.g.b.$'' indicates that the continuum background is negligible.}
\begin{ruledtabular}
\begin{tabular}{llllll}
  Mode  & $N^{obs}$ & $N^{obs}$  & Efficiency   & Branching Ratio        & Detector     \\
        &  (peak) & (continuum)  & (\%)         &     ($\times 10^{-3}$) &     \\ \hline
$\pip\pim\piz$
   &$1849852\pm 1360$&$31.0\pm 5.6$ &$38.13$
                                   &$21.37\pm 0.04^{+0.58}_{-0.56} $& BESIII~\cite{bes3pi2012} \\
   &$219691.0\pm 503.0 $ &   &$17.83 $    &$21.84\pm 0.05\pm 2.01$  & BESII~\cite{besj3pi04} \\
   &$166  $          &   &$2.68 $         &$15.0 \pm 2.0  $         & MARKII~\cite{Franklin83} \\
$\rpi$
   &$149.7  $        &   &$2.68 $         &$13.0 \pm 3.0  $         & MARKII~\cite{Franklin83} \\
   &$543.0\pm 105.6$ &   &$6.08$          &$10  \pm 2    $          & DESY-H.~\cite{wBartelg1976} \\
$\kstp\kam+c.c.$
   &$2285.0\pm 43.1 $&     &$5.814^{\ast}$&$4.57\pm 0.17\pm 0.70 $  & DM2~\cite{JJousset90jvp} \\
   &$24  $           &     &$2.13 $   &$2.6 \pm 0.8  $              & MARKII~\cite{Franklin83} \\
$\kstz\kazb+c.c$
   &$1192.0\pm 39.1 $&     &$3.500^{\ast}$&$3.96\pm 0.15\pm 0.60 $  & DM2~\cite{JJousset90jvp} \\
$\phi\eta$
   &$346.0\pm 21.2 $ &     &$6.286^{\ast}$&$0.64\pm 0.04\pm 0.11 $  & DM2~\cite{JJousset90jvp} \\
   &$2418.0\pm 65.2$ &     &$4.667^{\ast}$&$0.898\pm 0.024\pm 0.089$& BESII~\cite{besjfpaet05}\\
$\phi\etap$
   &$167.0\pm 13.5 $ &     &$4.736^{\ast}$&$0.41\pm 0.03\pm 0.08 $  & DM2~\cite{JJousset90jvp} \\
   &$728.0\pm 40.5 $ &     &$2.311^{\ast}$&$0.546\pm 0.031\pm 0.056$& BESII~\cite{besjfpaet05}\\
   &$31321\pm 201 $  &     &$4.690  $     &$0.510\pm 0.003\pm 0.032$& BESIII~\cite{bes16fetp} \\
$\omega\eta$
   &$378.0\pm 26.9 $ &     &$3.074^{\ast}$&$1.43\pm 0.10\pm 0.21 $  & DM2~\cite{JJousset90jvp} \\
   &$4927.0\pm 91.0$ &$n.g.b.$ &$3.631^{\ast}$&$2.352\pm 0.273  $   & BESII~\cite{besjopaet06} \\
$\omega\etap$
   &$6.0\pm 2.5 $ &  & $0.388^{\ast}$&$0.18^{+0.10}_{-0.08}\pm 0.03 $& DM2~\cite{JJousset90jvp} \\
   &$218.0\pm 32.8 $ &$n.g.b.$&$1.672^{\ast}$&$0.226\pm 0.043  $     & BESII~\cite{besjopaet06} \\
   &$137\pm 20 $     &$n.g.b.$&$0.050 $   &$0.208\pm 0.030\pm 0.014$ & BESIII~\cite{besretp2017} \\
$\roz\eta$
   &$299.0\pm 34.0 $ &     &$17.921^{\ast}$&$0.194\pm 0.017\pm 0.029$& DM2~\cite{JJousset90jvp} \\
$\roz\etap$
   &$19.2\pm 7.5 $  &     &$2.690^{\ast}$&$0.083\pm 0.030\pm 0.012 $ & DM2~\cite{JJousset90jvp} \\
   &$3621\pm 83 $ &$57.13\pm 11.03$ &$3.381 $&$0.0790\pm 0.0019\pm 0.0049$& BESIII~\cite{besretp2017} \\
$\phi\piz$
   &$24         $ &    &$8.08 $       &$<6.4\times 10^{-3}~(90\% C.L.)$& BESII~\cite{besjfpaet05}\\
&($838.5\pm 45.8$)\footnote{
\begin{minipage}{12.3cm}\mbox{}The solution of constructive interference between $\jpsi \to \phi\piz$ and $\jpsi \to K^+K^- \piz$ decays.  \end{minipage} }
                  &$n.g.b.$&$21.79 $ &$(2.94\pm 0.16 \pm 0.16)\times 10^{-3} $ & BESIII~\cite{bes15fpi}\\
&($35.3\pm 9.3 $)\footnote{
\begin{minipage}{12.2cm}\mbox{}The solution of destructive interference between $\jpsi \to \phi\piz$ and $\jpsi \to K^+K^- \piz$ decays.  \end{minipage} }
                  &$n.g.b.$&$21.79 $ &$(1.24\pm 0.33 \pm 0.30)\times 10^{-4} $ & BESIII~\cite{bes15fpi}\\
$\omega\piz$
   &$222.0\pm 19.0 $ &     &$7.171^{\ast}$&$0.360\pm 0.028\pm 0.054$  & DM2~\cite{JJousset90jvp} \\
   &$2090.0\pm 67.3$ &$6.88\pm 2.85$ &$6.57$&$0.538\pm 0.012\pm 0.065$& BESII~\cite{besjopaet06}
\end{tabular}
\end{ruledtabular}
\end{table*}

\subsection{$\jpsi \to VP$ decay }
\begin{table*}[bth]
\caption{\label{tab_jpscmp}Comparison of various fitting results for the destructive interference solution data and positive angle case. The data with star (${\star}$) exclude the measurement of $\psp \to \omega\eta$ decay from BESII~\cite{besjopaet06}. $n_d$ indicates the degree of freedom.}
\begin{ruledtabular}
\begin{tabular}{llllll}
$\vec{\eta}$& All             & All           & DM2            &  DM2 \& BESII  &  DM2 \& BESIII   \\
\& $\chi^2$ & data            & data$^{\star}$& data           &data$^{\star}$  &  data      \\ \hline
$\chi^2$    &$4302.444$       &$1962.792$      &$91.254$       &$345.825$      &$239.325$ \\
$\chi^2/n_d$&$330.6$          &$163.6$         &$45.63$        &$49.40$        &$29.92$  \\  \hline
$  \phi  $  &$(71.42\pm 1.82)^\circ$ &$(64.47\pm 1.91)^\circ$
                       &$(70.92\pm 3.43)^\circ$ &$(61.60\pm 3.46)^\circ$ &$(69.52\pm 3.18)^\circ$ \\
$    A   $  &$ 2.697\pm 0.011$ &$2.631\pm 0.011$&$2.013\pm 0.209$&$2.460\pm 0.033$&$2.914\pm 0.018$ \\
$    D   $  &$ 0.538\pm 0.006$ &$0.562\pm 0.006$&$0.395\pm 0.044$&$0.497\pm 0.010$&$0.690\pm 0.017$ \\
$   \Dp  $  &$-0.489\pm 0.017$ &$-0.638\pm0.018$&$-0.223\pm0.049$&$-0.679\pm0.032$&$0.171\pm 0.069$ \\
$   s_P  $  &$-0.341\pm 0.004$ &$-0.472\pm0.005$&$-0.530\pm0.046$&$-0.755\pm0.010$&$-0.208\pm0.028$ \\
$\theta_P$  &$-19.96^\circ   $ &$-28.17^\circ$  &$-32.76^\circ$  &$-48.98^\circ$  &$-11.98^\circ$    \\
$   s_V  $  &$ 0.557\pm 0.002$ &$0.557\pm 0.001$&$0.308\pm 0.045$&$0.221\pm 0.010$&$0.657\pm 0.025$  \\
$\theta_V$  &$ 33.82^\circ   $ &$33.85^\circ$   &$17.94^\circ$   &$12.76^\circ$   &$41.10^\circ$ \\ \hline
$f_{dm2}$   &$ 0.590\pm 0.010$&$0.603\pm 0.011$&$1.020\pm 0.212$&$0.698\pm 0.022$ &$0.422\pm 0.014$ \\
$f_{bes2}$  &$ 1.177\pm 0.010$&$1.166\pm 0.010$&                &$1.295\pm 0.034$ &                \\
$f_{bes3a}$ &$ 0.579\pm 0.009$&$0.686\pm 0.015$&                &                 &$0.281\pm 0.012$ \\
$f_{bes3b}$ &$ 1.233\pm 0.010$&$1.864\pm 0.01-$&                &                 &$1.326\pm 0.011$
\end{tabular}
\end{ruledtabular}
\end{table*}

For $\jpsi$ decaying to ${\mathbf V} {\mathbf P}$ final state, there are lots of experimental results, some of which are summarized in Table~\ref{tab_jpstovp}. However, a few of measurements are obtained forty years ago with low statistic samples. In Refs.~\cite{bJeanMarie76} and~\cite{gAlexander1978}, the branching fraction of $\rpi$ is measured to be $(1.3 \pm 0.3)\times 10^{-2}$ and $(1.6 \pm 0.4)\times 10^{-2}$ based on 50 and 84 thousand $\jpsi$ events, respectively.
In Refs.~\cite{Franklin83} and~\cite{wBartelg1976}, the branching fraction of $\rpi$ is measured to be $(1.3 \pm 0.3)\times 10^{-2}$ and $(1.0 \pm 0.2)\times 10^{-2}$ based on 0.4 and 0.87 million $\jpsi$ events, respectively.
In Ref.~\cite{Brandelik:1978}, the study is mainly focused on $\jpsi$ radative decay into $\pi\pi\gamma$ and $KK\gamma$. With 1.71 million $\jpsi$ events, the branching fraction of $\rpi$ is merely obtained as a consistency check. These kinds of measurements involving $\rpi$ final state are excluded from the following analysis. The measurements of $\jpsi \to \kstp\kam+c.c.$ from Ref.~\cite{Franklin83},
$\jpsi \to \phi\piz$ from Ref.~\cite{besjfpaet05}, $\jpsi \to \rho^{\pm}\pi^{\mp},~K^{*\pm}K^{\mp}$ from Ref.~\cite{wBraunschweig1976}, and $\jpsi \to \kstp\kam+c.c.,~\kstz\kazb+c.c$ from Ref.~\cite{fVannucci77} are not adopted either due to low statistic.

Both Ref.~\cite{mk3jvp85} and Ref.~\cite{mk3jvp88} made a systematical measurement for $\jpsi$ decaying into ${\mathbf V} {\mathbf P}$ final state. The latter data, consisting of $5.8 \times 10^{6}$ produced $\jpsi$'s, represents a twofold increase over the former data ($2.7 \times 10^{6}$), therefore the better accuracy is realized in the latter analysis. Although all ${\mathbf V} {\mathbf P}$ channels, viz., $\rpi$, $\roz\eta$, $\roz\etap$, $\omega\piz$, $\omega\eta$, $\omega\etap$, $\phi\piz$, $\phi\eta$, $\phi\etap$, $\kstp\kam+c.c.$, and $\kstz\kazb+c.c$, are measured, but the number of events and the corresponding efficiencies are not provided, which leads to impossibility to include these results in this analysis.

In addition, in Ref.~\cite{bes3pi96}, cross sections of $\rpi$ final state are measured at 29 different energy points covering a 40 MeV interval spanning the $\jpsi$ resonance. Based on this data sample, which corresponds to a total integrated luminosity of 238 nb$^{-1}$, the branching fraction is determined to be $(1.21\pm 0.20)$\%. Such information is too distinctive to be combined with those in Table~\ref{tab_jpstovp}.
In Ref.~\cite{bes09ksk} resorting to partial wave analysis technique measured is the branching fraction of $\jpsi \to  \kstp\kam+c.c.$, whose feature is too different to be merged with other information.
In both Ref.~\cite{aAubert3pi04} and Ref.~\cite{aAubert07oget} the initial state radiation technique is used to obtain the branching fractions of $\jpsi \to \pip\pim\piz$ and $\jpsi \to \omega\eta$. Such kinds of results can not be utilized in the present scheme of analysis.

Based on a sample of 1.31 billion $\jpsi$ events~\cite{bes15fpi}, the $K^+K^-$ mass spectrum is scrutinized and observed is a clear structure due to the interference between $\jpsi \to \phi\piz$ and $\jpsi \to K^+K^- \piz$ decays. Such a interference yields two possible solutions of branching fraction, that is $(2.94\pm 0.16 \pm 0.16)\times 10^{-6}$ and $(1.24\pm 0.33 \pm 0.30)\times 10^{-7}$, which correspond to the  constructive and destructive interferences, respectively. In the following fit, two results will be dealt with separately.

The minimization estimator for $\jpsi$ is similar to that of $\psp$ as defined in Eq.~\eref{chisqbb} and the fit yields
\beq
\begin{array}{rcll}
  \phi   &=&-66.52^\circ\pm 1.82^\circ~, &\mbox{ or } ~+71.42^\circ\pm 1.82^\circ~, \\
    A    &=&~2.622 \pm 0.011~, &\mbox{ or }~2.697 \pm 0.011~,\\
    D    &=&~0.523 \pm 0.006~, &\mbox{ or }~0.538 \pm 0.006~,\\
  \Dp    &=&-0.475 \pm 0.017~, &\mbox{ or }-0.489 \pm 0.017~,\\
  s_P    &=&-0.341 \pm 0.004~, &\mbox{ or }-0.341 \pm 0.004~,\\
\theta_P &=&-20.17^\circ~ , &\mbox{ or } -19.96^\circ~ ,\\
  s_V    &=&~0.557 \pm 0.002~,&\mbox{ or } ~0.557 \pm 0.002~,\\
\theta_V &=&~33.86^\circ~, &\mbox{ or } ~33.82^\circ~,\\
  f_{dm2}&=&~0.624 \pm 0.011~,&\mbox{ or } ~0.590 \pm 0.010~,\\
 f_{bes2}&=&~1.246 \pm 0.010~,&\mbox{ or } ~1.177 \pm 0.010~,\\
f_{bes3a}&=&~0.613 \pm 0.010~,&\mbox{ or } ~0.579 \pm 0.009~,\\
f_{bes3b}&=&~1.305 \pm 0.010~,&\mbox{ or } ~1.233 \pm 0.010~;
\end{array}
\label{fitjpa}
\eeq
with $\chi^2=4302$ for the destructive interference solution, and
\beq
\begin{array}{rcll}
  \phi   &=&-66.92^\circ\pm 1.77^\circ~, &\mbox{ or } ~+71.82^\circ\pm 1.77^\circ~, \\
    A    &=&~2.633 \pm 0.011~, &\mbox{ or }~2.713 \pm 0.011~,\\
    D    &=&~0.536 \pm 0.006~, &\mbox{ or }~0.552 \pm 0.006~,\\
  \Dp    &=&-0.501 \pm 0.017~, &\mbox{ or }-0.516 \pm 0.018~,\\
  s_P    &=&-0.397 \pm 0.004~, &\mbox{ or }-0.397 \pm 0.004~,\\
\theta_P &=&-23.38^\circ~ , &\mbox{ or } -23.38^\circ~ ,\\
  s_V    &=&~0.492 \pm 0.002~,&\mbox{ or } ~0.492 \pm 0.002~,\\
\theta_V &=&~29.47^\circ~, &\mbox{ or } ~29.47^\circ~,\\
  f_{dm2}&=&~0.620 \pm 0.011~,&\mbox{ or } ~0.584 \pm 0.010~,\\
 f_{bes2}&=&~1.224 \pm 0.010~,&\mbox{ or } ~1.152 \pm 0.010~,\\
f_{bes3a}&=&~0.664 \pm 0.011~,&\mbox{ or } ~0.625 \pm 0.010~,\\
f_{bes3b}&=&~1.282 \pm 0.010~,&\mbox{ or } ~1.207 \pm 0.010~;
\end{array}
\label{fitjpb}
\eeq
with $\chi^2=4200$ for the constructive interference solution, where $\theta_\alpha = \arcsin s_\alpha~(\alpha=P,V)$.

From a pure viewpoint of hypothesis test~\cite{IGHuges2010,AGFrodeson19790}, the ratio of the chi square value to the number of degrees of freedom should approximate one for a good fit, but the values of $\chi^2$ for fitting results in Eqs.~\eref{fitjpa} and \eref{fitjpb} is horrendously large.
Since the accuracy of $\jpsi$ data is generally higher than that of $\psp$, the discrepancies between the different experiments become much more prominent. The data summarized in Table~\ref{tab_vpdata} can be grouped into four sets: two from BESIII, one with total luminosity 394.65 pb$^{-1}$, the other with luminosity 79.63 pb$^{-1}$; one from BESII with luminosity 15.89 pb$^{-1}$ and one from MD2, with total luminosity 5.053 pb$^{-1}$. Four normalization factors of luminosity are introduced to alleviate the possible inconsistence among the data from different experiment groups. Nevertheless, such a kind of treatment is obviously not sufficient enough so that certain great deviations still exist which lead to huge chisquare  value.

In order to figure out the effect due to the discrepancy of different experiments on the value of $\chi^2$, the fits for different datasets are performed respectively, viz., DM2 data, DM2 and BESII data, DM2 and BESIII data, and all dataset, the results are tabulated in Table~\ref{tab_jpscmp}. The value of $\chi^2$ is minimum for a sole experiment dataset (DM2 data), and the values increase when more experiment datasets are fit together (DM2 and BESII datasets or DM2 and BESIII datasets). It can be seen when all data are fit together, the value of $\chi^2/n_d$ enhances almost one order of magnitude. Moreover, after some filtration trial it is found that the inconsistence of the measurement of $\psp \to \omega\eta$ decay from BESII~\cite{besjopaet06} from the other measurements is rather obviously. The fit is performed for all data except for this one, the results are presented in the second column of Table~\ref{tab_jpscmp}. The comparison of chisquares of the first two columns indicates that the sole measurement of $\omega\eta$ channel from BESII contributes more than half of chi square value, which seriously deteriorates the $\chi^2$ probability of the fit. If we compare the two branching fractions from DM2~\cite{JJousset90jvp} and BESII~\cite{besjopaet06}, that is $(1.43\pm 0.10\pm 0.21)\times 10^{-3}$ and $(2.352\pm 0.273)\times 10^{-3}$, the former is only about the half of the latter. The great discrepancy renders huge chi square, which can only be settled down by further more accurate experimental measurement.

Of course, there is a factor that has been neglected for data analysis. In the fitting solely considered are the statistic uncertainties, if the systematic uncertainties are included as well, it is expected that the chisquare could be decreased to one half or one third of the present value. As a matter of fact, we also performed a fit with increased error of $\omega\eta$ channel. If $4927.0\pm 571.0$ instead of $4927.0\pm 91.0$ is used, $\chi^2=2055$ instead of $\chi^2=4302$. Anyway, even so the value of $\chi^2$ is still too large to be satisfied from a point of statistical view.

It also exists the possibility that the present parametrization form is not exquisite enough to describe all data perfectly, but only more precise and consistent experimental data can furnish quantitative evidence for or against the present phenomenology model and pin down the problem we come across here.
The absence of a set of ideally experimental data is keenly felt.

Last but not least, we also perform the fit for data related to $\roz\eta$, $\roz\etap$, $\omega\piz$, and $\phi\piz$ final states, the fitting yields a $\chi^2$ of 3.91 with the number of degrees of freedom being 3 for the destructive interference case~:
\beq
\begin{array}{rcl}
      D   &=&~0.404\pm 0.037~, \\
    s_P   &=&-0.288\pm 0.068~, \\
  \theta_P&=&-16.73^\circ~ , \\
    s_V   &=&~0.595\pm 0.003~, \\
  \theta_V&=&~36.50^\circ~, \\
  f_{dm2} &=&~1.096 \pm 0.248~,\\
 f_{bes2} &=&~1.773 \pm 0.329~;
\end{array}
\label{ftjvpas}
\eeq
and a $\chi^2$ of 3.97 with the number of degrees of freedom being 3 for the constructive interference case~:
\beq
\begin{array}{rcl}
      D   &=&~0.403\pm 0.036~, \\
    s_P   &=&-0.285\pm 0.068~, \\
  \theta_P&=&-16.57^\circ~ , \\
    s_V   &=&~0.660\pm 0.010~, \\
  \theta_V&=&~41.30^\circ~, \\
  f_{dm2} &=&~1.112 \pm 0.252~,\\
 f_{bes2} &=&~1.805 \pm 0.337~.
\end{array}
\label{ftjvpbs}
\eeq
For the destructive interference case the fit values of $\theta_P$ and $\theta_V$ are marginally consist with those of overall fit, which means that these four channels are necessary, sufficient, and efficient for measurement of the mixing angles of pseudoscalar and vector mesons.

\begin{table}[bth]
\caption{\label{tab_brdt}Branching ratios of $\psp,\jpsi \to V P $ extracted from PDG2022~\cite{pdg2022}. $Q_h$ is defined in Eq.~\eref{qcdrule} and calculated by the ratio of $B(\psp \to VP)$ to $B(\jpsi \to VP)$. $(\kstp\kam)_{c.c.}$ and $(\kstz\kazb)_{c.c}$ indicate $\kstp\kam+c.c.$ and $\kstz\kazb+c.c.$ respectively.
 }
\begin{ruledtabular}
\begin{tabular}{llll}
  Mode  & $B(\psp \to VP)$      &$B(\jpsi \to VP)$     & $Q_h$\\
        & ($\times 10^{-5}$)    &($\times 10^{-3}$)    & (\%) \\ \hline
$\rpi$           &$  3.2 \pm 1.2 $ &$ 16.9  \pm 1.5   $ &$ 0.2\pm  0.1  $\\
$(\kstp\kam)_{c.c.}$ &$  2.9 \pm 0.4 $ &$ 6.0   \pm 1.0   $ &$ 1.1\pm  0.2  $\\
$(\kstz\kazb)_{c.c}$ &$ 10.9 \pm 2.0 $ &$ 4.2   \pm 0.4   $ &$ 2.6\pm  0.5  $\\
$\phi\eta$       &$ 3.10 \pm 0.31$ &$ 0.74  \pm 0.08  $ &$ 4.2\pm  0.6  $\\
$\phi\etap$      &$ 1.54 \pm 0.20$ &$ 0.46  \pm 0.05  $ &$ 3.3\pm  0.6  $\\
$\omega\eta$     &$   <1.1       $ &$ 1.74  \pm 0.20  $ &$ <0.632       $\\
$\omega\etap$    &$ 3.2  \pm 2.5 $ &$ 0.189 \pm 0.018 $ &$16.9\pm  13.3 $\\
$\roz\eta$       &$ 2.2  \pm 0.6 $ &$ 0.193 \pm 0.023 $ &$11.4\pm  3.4  $\\
$\roz\etap$      &$ 1.87 \pm 1.67$ &$ 0.081 \pm 0.008 $ &$23.1\pm  20.7 $\\
$\omega\piz$     &$   <0.004     $ &$ 0.45  \pm 0.05  $ &$ <0.0089      $\\
$\phi\piz$       &$ 2.1  \pm 0.6 $ &$ (2.94 \pm 0.23)\cdot 10^{-3}$\footnote{\begin{minipage}{3.50cm}\mbox{}The constructive solution.\end{minipage} }
& $(7.1\pm 2.1) \cdot 10^{2}$\\
                 &$              $ &$ (1.24\pm 0.45)\cdot
10^{-4}$\footnote{\begin{minipage}{3.40cm}\mbox{}The destructive solution. \end{minipage} }
& $(1.7\pm 0.8) \cdot 10^{4}$ \\
\end{tabular}
\end{ruledtabular}
\end{table}

\begin{table*}[bth]
\caption{\label{tab_brdtft}Fit results of branching ratio method for $\psp,\jpsi \to V P$ decay.
$n_{d.o.f}$ indicates degree of freedom.}
\begin{ruledtabular}
\begin{tabular}{lllll}
$\vec{\eta}$& $\psp \to VP$   & \multicolumn{3}{c}{$\jpsi \to VP$}     \\
\& $\chi^2$ &                 & \multicolumn{2}{c}{constructive solution}
                                                       & destructive solution      \\ \hline
$\chi^2$    &   1.995          &  184.075          &  184.075          &  169.651          \\
$n_{d.o.f}$ &   3              &  5                &  5                &  5               \\ \hline
$  \phi  $  &$(-133.56\pm 13.54)^\circ$ &$( 61.65\pm 31.87)^\circ$
                      &$(-61.65\pm 20.89)^\circ $ &$(-152.18\pm 44.56)^\circ$ \\
$    A   $  &$ 0.887 \pm 0.092$ &$ 0.796 \pm 0.027$ &$ 0.796 \pm 0.031$ &$ 0.714 \pm 0.040$ \\
$    D   $  &$ 0.618 \pm 0.079$ &$-0.192 \pm 0.008$ &$-0.192 \pm 0.008$ &$ 0.202 \pm 0.007$ \\
$   \Dp  $  &$ 0.247 \pm 0.166$ &$-0.277 \pm 0.088$ &$-0.277 \pm 0.060$ &$-0.233 \pm 0.071$ \\
$   s_P  $  &$-0.227 \pm 0.074$ &$-0.585 \pm 0.018$ &$-0.585 \pm 0.022$ &$-0.439 \pm 0.051$ \\
$\theta_P$  &$-13.10^\circ    $ &$-35.84^\circ    $ &$-35.84^\circ    $ &$-26.04^\circ    $ \\
$   s_V  $  &$ 0.029 \pm 0.141$ &$ 0.493 \pm 0.002$ &$ 0.493 \pm 0.005$ &$ 0.594 \pm 0.003$ \\
$\theta_V$  &$ 1.64^\circ    $ &$ 29.52^\circ    $ &$ 29.52^\circ    $ &$ 36.44^\circ    $ \\
\end{tabular}
\end{ruledtabular}
\end{table*}

\section{Discussion}\label{xct_dskn}
Besides the cross section approach (CSA) used to analyze the data, there is another way to deal with the information involving $\psp,\jpsi \to {\mathbf V} {\mathbf P}$ decay, which is call branching ratio method (BRM). The idea is fairly simple, the so called ``reduced branching ratio'' is related to the square of amplitude directly~\cite{Haber,a11}, that is
\beq
\tilde{B} (\psi \to f) = |X_f+Y_f|^2~,
\label{rbrwamp}
\eeq
where $X_f$ and $Y_f$ are defined in Eqs.~\eref{defx} and \eref{defy} for a certain ${\mathbf V} {\mathbf P}$   final state $f$, and the reduced branching ratio is defined as follows
\beq
\tilde{B} (\psi \to f) = \frac{{B}(\psi \to f)}{q_f^3}~,
\label{rbrdef}
\eeq
where ${B}(\psi \to f)$ is the branching ratio of $\psi (\psi=\jpsi, \psp)$ decays to the final state $f$, and $q_f$ defined in Eq.~\eref{qfexprn}, is the momentum of either particle in the center of mass system for two-body decay. According to the formula~\eref{rbrwamp} together with the information in  Table~\ref{tab_brdt}, the fits are performed and the fitting results are displayed in Table~\ref{tab_brdtft}.

Comparing results of two kinds of fits, for $\psp$ case, the values of $\chi^2$ are both reasonable, merely one solution is found. Some results are similar such as $A, D, s_P$ while others are different
such as $\Dp, s_V$. For $\jpsi$ case, the similarity is much worse, especially for the destruction solution, only one minimum instead of two is found for the phase angle. As a matter of fact, since the branching ratios herein are averaged results that combine the various measurements due to many experiments, such an admixture blurs the discrepancy between distinctive experimental analysis. Therefore, the following discussion is based on the results due to CSA instead of BRM.

\begin{table}[hbt]
\caption{\label{fitifnftbd}The fit results of phase angle for $\psp$ and $\jpsi$ two-body decays, some of which are from Ref.~\cite{moxh2024}. The subscripts $d$ and $c$ indicate the results for destructive and constructive cases respectively. }
\center
\begin{tabular}{lll}\hline \hline
Decay mode  & \multicolumn{2}{c}{phase angle $\phi$ (in degree)} \\ \hline
$\psp \to B_{8}  \overline{B}_{8} $& $-94.59\pm 1.31$  & $+85.42\pm 2.25$     \\
$\jpsi\to B_{8}  \overline{B}_{8} $& $-84.81\pm 0.70$  & $+95.19\pm 0.70$     \\
$\psp \to B_{10} \overline{B}_{10}$& $-75.51\pm 4.87$  & $+104.49\pm 4.91$     \\
$\jpsi\to B_{10} \overline{B}_{10}$& $-96.28\pm 17.23$ & $+83.27\pm 11.38$     \\
$\jpsi\to B_{10} \overline{B}_{8} $& $-89.97\pm 37.17$ & $+101.20\pm 71.87$     \\
$\psp \to PP                      $& $-58.19\pm 5.47$  & $+92.82\pm 5.62$    \\
$\jpsi\to PP                      $& $-87.25\pm 8.60$  & $+92.14\pm 8.61$     \\
$\psp \to VP                      $& $(-131.55\pm 13.05)_d$  &     \\
                                   & $(-144.31\pm 20.93)_c$  &     \\
$\jpsi\to VP                      $& $(-66.52\pm 1.82)_d$  & $(+71.42\pm 1.82)_d$     \\
                                   & $(-66.92\pm 1.77)_c$  & $(+71.42\pm 1.77)_c$     \\
\hline \hline
\end{tabular}
\end{table}

Table~\ref{fitifnftbd} summarizes various fitting results of phase angle for $\psp$ and $\jpsi$ two-body decays. The most exceptional solution is due to $\psp \to {\mathbf V} {\mathbf P}$ decay. Such a situation can not help reminding of the famous ``$\RP$ puzzle'' in charmonium decays.

Theoretically, the OZI (Okubo-Zweig-Iizuka)~\cite{ozi} suppressed decays of $\jpsi$ and $\psp$ to hadrons
are via three gluons or a photon, in either case, the perturbative QCD (pQCD) provides a relation~\cite{appelquist}
\begin{eqnarray}
Q_h &=&\frac{{\cal B}_{\psp \ra h}}{{\cal B}_{\jpsi \ra h}}
=\frac{{\cal B}_{\psp \ra \EE}}{{\cal B}_{\jpsi \ra \EE}}
\approx (13.28\pm 0.29)\%~,
\label{qcdrule}
\end{eqnarray}
where ${\cal B}_{\psp \ra \EE}$ and ${\cal B}_{\jpsi \ra \EE}$ are taken from PDG2022~\cite{pdg2022}.
This relation is expected to be held to a reasonable good degree for both inclusive and exclusive decays. The so-called ``$\RP$ puzzle'' is that the prediction by Eq.~\eref{qcdrule} is severely violated in the $\RP$ and several other decay channels. The first evidence for this effect was found by Mark-II Collaboration in 1983~\cite{Franklin83}. From then on many theoretical explanations have been put forth to decipher this puzzle, refer to the treatise~\cite{moxh07rprv} for a detailed review.

From the theoretical point of view, since the $Q$-value is smaller than the expectation for $\rpi$, it may be caused either by enhanced $\jpsi$ or suppressed $\psp$ decay rate. Another possibility is by both. Then the relevant theoretical speculations can be classified into three categories~:
\begin{enumerate}
\item $\jpsi$-enhancement hypothesis, which attributes the small $Q$-value to the enhanced branching fraction of $\jpsi$ decays~\cite{brodsky81}-\cite{brokar}.
\item $\psp$-suppress hypothesis, which attributes the small $Q$-value to the suppressed branching fraction of $\psp$ decays~\cite{chen}-\cite{rosnersd}.
\item Other hypotheses, which are not included in the above two categories~\cite{lixq}-\cite{hoyer}.
\end{enumerate}
With more and more experimental data are obtained, many theoretical explanations have been ruled out and some still need to be tested. However, ``$\RP$ puzzle'' seems still a puzzle since no propose can explain all existing experimental data satisfactorily and naturally. If we scrutinize the $Q_h$ values in Table~\ref{tab_brdt}, it can seen that many values are suppressed relative to the expected value 13.28\% while some are enhanced, {\em a fortiori} for $\phi\piz$ channel, the $Q_h$ value is several orders of magnitude greater than the expectation. Furthermore the deviation seems rather arbitrary, no regularity can be found. At the same time, if we investigate Table~\ref{fitifnftbd}, it can be seen that the phase angle for $\psp \to {\mathbf V} {\mathbf P}$ decay is rather abnormal from the other decays. So more we think about various information, more problems spring up. It seems to trigger a Pandora's Box of questions~:
\begin{enumerate}
\item $\psp$ and $\jpsi$ are both $S$-state of chamonium, the decay pattern is similar for some channels but rather different for the others, the prominent example is about $\rpi$ final state, some studies of which have been performed in Ref.~\cite{besrp2005}. What is the reason for such grotesquery situations?
\item Is the abnormal feature only for ${\mathbf V} {\mathbf P}$ mode, or for all $VP$-like modes, such as $SV$, $SA^-$, $PA^-$, $VT$, $VA^+$, $TA^-$, and $A^+ A^-$ ? ($S, P, V, A$, and $T$ denote scalar, pseudoscalar, vector, axial vector, and tensor, respectively; refer to Table~\ref{octetmesons} and Table~\ref{termalwineffh} for more details. )
\item Does the exceptional phase angle have a connection with the abnormal $Q_h$ value? Is there a profound rationale for such a connection?
\item From the viewpoint of phase angle and by virtue of fit result, the baryonic mode seems more normal than the mesonic mode. What is the reason for such a difference?
\item From the viewpoint of phase angle and by virtue of fit result, $\jpsi$ decay seems more normal than $\psp$. What is the reason for such a difference?
\item Does such a bizarre situation also exist for bottomonium decay?
\item There is a suggestion that $\psp$ is not a pure state but a mixing of $\psi (2^3 S_1)$ state and $\psi (1^3 D_1)$. Such a opinion is adopted to explain the decrease $Q_h$ value for $\rhopi$ channel~\cite{rosnersd} and the increase $Q_h$ value for $\kskl$ channel~\cite{psippkskl}. Anyway, could this suggestion explain decay behaviors for all kinds of modes?
\item Glueball was once put forth to explain ``$\RP$ puzzle'' ~\cite{freund}-\cite{anselm}, but the experimental research is unfavorable to the detailed analyses and deductions due to this kind of scenario. However, the eccentric decay pattern of ${\mathbf V} {\mathbf P}$ mode shown by their various $Q_h$ values indeed implies the peculiar feature of this mode. The doubt is aggravated by the fitted mixing angles of $\theta_P$ and $\theta_V$, none of which is consist with theoretical expectation. According to mass matrix analysis, $\theta_P=-11.3^\circ$ or $-24.5^\circ$ and $\theta_V=39.2^\circ$ or $36.5^\circ$ corresponding to the quadratic or linear mass assumptions respectively. The deviation from expectation maybe indicate the admixture of pseudoscalar meson with some glueball-like component. This is an issue need to be studied further.
\end{enumerate}

It is true that a lot of measurements are available now, but even more precise and systematical measurements are needed to clarify the dynamics of charmonium decay.

\section{Summary}\label{xct_sum}
The flavor-singlet principle, with assuming the flavor symmetry breaking effects (both strong and electromagnetic breaking effects) as a special $SU(3)$ octet, furnishes a criterion to figure out the effective interaction Hamiltonian in tensor form for all kinds of two-body final states decaying from a charmonium resonance.
The generalized inherent ${C}$-parity for multiplet is introduced, which plays a crucial role for determining the form of effective Hamiltonian, especially for mesonic final states.
Resorting to the nonet notion, both octet and singlet representations for meson description are synthesized together to acquire the effective Hamiltonian in a concise way, then solving the meson mixing
problem in charmonium decay. As an application, by virtue of this scenario the relative phase between the strong and electromagnetic amplitudes is measured for vector-pseudoscalar meson final state, more information involving the interaction coupling coefficients is obtained, all of which deepen our understanding of the dynamics of charmonium decay. Furthermore, the fit results indicate that the measurements of four final states, that is $\roz\eta$, $\roz\etap$, $\omega\piz$, and $\phi\piz$, are necessary, sufficient, and efficient for the study of mixing angles of pseudoscalar and vector multiplets.

In analysis of data samples taken in $\EE$ collider, the details of experimental effects, such as energy spread and initial state radiative correction are taken into consideration in order to make full advantage of experimental information and acquire the comprehensive results. However, the discrepancy between different experimental measurement leads to large chi square, which is fairly unfavorable from the statistic viewpoint. The data analysis of this paper makes it urgent that further more precisely and systematically experimental measurements should be performed based on BESIII colossal data sample of charmonium decay, in order to figure out the unclear issues we come across here.

By virtue of present analysis, the uniform parametrization scheme provides a general description for various kinds of charmonium two-body decays and lays an extensive foundation for more profound dynamics exploration in the future.

\section{Appendix}
This appendix is devoted to two issues. The first one is about the ${C}$-parity transformation of meson octet.

For two meson octets, denoting respectively by $O_1$ and $O_2$, defined are the following terms which may be allowed or forbidden in the effective Hamiltonian:
\beq
[O_1 O_2]_0 = (O_1)^i_j (O_2)^j_i~~,
\eeq
\beq
([O_1 O_2]_f )^i_j = (O_1)^i_k (O_2)^k_j -(O_1)^k_j (O_2)^i_k~~,
\eeq
and
\beq
([O_1 O_2]_d )^i_j = (O_1)^i_k (O_2)^k_j +(O_1)^k_j (O_2)^i_k
-\frac{2}{3} \delta^i_j \cdot (O_1)^i_j (O_2)^j_i~~.
\eeq
The aforementioned generalized inherent ${C}$-parity for meson octet is introduced as a criterion for Hamiltonian terms, and its value (denoted as $\eta_O$ ) is set to be equal to that of the neutral particle of the corresponding octet. In Table~\ref{octetmesons} listed are some observed light mesons that are classified into distinctive octets. The value of generalized ${C}$-parity equals to the ${C}$-parity presented in the table.

\begin{table}[hbt]
\caption{\label{octetmesons} Some meson octet particles. For $I=0$ meson, the mixing between singlet and octet  always exists. The $f^{\prime}$ and $f$ are mixing states as defined in Eq.~\eref{mixmsnffp} or in Eqs.~\eref{defoffpp} and ~\eref{defoffp}. In addition, $K_{1A}$ and $K_{1B}$ are nearly equal ($45^{\circ}$) mixtures of the $K_1(1270)$ and $K_1(1400)$. } \center
\begin{tabular}{cc|cccc}\hline \hline
 Octet & $J^{PC}$  & $I=1$       &  $I=1/2$    & $I=0$  & $I=0$  \\
       & & $u\bar{d},\bar{u}d,$  & $u\bar{s},d\bar{s};$    & $f^{\prime}$  & $f$  \\
       & & $(d\bar{d}-u\bar{u})/\sqrt{2}$  &  $\bar{d}s,-\bar{u}s$  &  &   \\ \hline
 $P$   & $0^{-+}$ & $\pi$       &  $K$        & $\eta$ & $\etap(958)$   \\
 $V$   & $1^{--}$ & $\rho(770)$  & $\kst(892)$   & $\phi(1020)$ & $\omega(782)$   \\
 $A^-$ & $1^{+-}$ & $b_1(1235)$  & $K_{1B}$      & $h_1(1380)$  & $h_1(1170)$   \\
 $S$   & $0^{++}$ & $a_0(1450)$  & $\kst_0(1430)$& $f_0(1710)$  & $f_0(1370)$   \\
 $A^+$ & $1^{++}$ & $a_1(1260)$  & $K_{1A}$      & $f_1(1420)$  & $f_1(1285)$   \\
 $T$   & $2^{++}$ & $a_2(1320)$  & $\kst_2(1430)$& $f^{\prime}_2(1525)$
                                                      & $f_2(1270)$   \\
\hline \hline
\end{tabular}
\end{table}

\begin{table}[hbt]
\caption{\label{termalwineffh} The determination of interaction terms in the effective Hamiltonian. The symbol $[O_1 O_2]_{x}$ is shorthand of $[O_1 O_2]_{0},[O_1 O_2]_{d}$, and $[O_1 O_2]_{f}$. Herein there are essentially two types of Hamiltonian forms, that is ``$yyn$'' which means both $[O_1 O_2]_{0}$ and $[O_1 O_2]_{d}$ terms are allowed in the effective Hamiltonian, and ``$nny$'' which means only $[O_1 O_2]_{f}$ term is allowed. The symbol ``$y$'' indicates the allowed term, while ``$n$'' indicates forbidden. $O_1$ and $O_2$ denote octets $S,P,V,T, A^+$, and $A^-$, which are shown in Table~\ref{octetmesons}. The superscript of symbol indicates the generalized inherent ${C}$-parity of corresponding octet.} \center
\begin{tabular}{c|cccccc}\hline \hline
$[O_1 O_2]_{x}$& $S^+$ & $P^+$ & $V^-$ & $T^+$ & $A^+$  & $A^-$ \\ \hline
      $S^+$  &   $nny$ & $nny$ & $yyn$ & $nny$ & $nny$  & $yyn$ \\
      $P^+$  &         & $nny$ & $yyn$ & $nny$ & $nny$  & $yyn$ \\
      $V^-$  &         &       & $nny$ & $yyn$ & $yyn$  & $nny$ \\
      $T^+$  &         &       &       & $nny$ & $nny$  & $yyn$ \\
      $A^+$  &         &       &       &       & $nny$  & $yyn$ \\
      $A^-$  &         &       &       &       &        & $nny$  \\
 \hline \hline
\end{tabular}
\end{table}

\begin{table}[hbt]
\caption{\label{ptmsnform}Amplitude parametrization form for decays of the $\psp$ or $\jpsi$ into PT final states. The table can also be used for other similar decays by appropriate change in labeling.
} \center
\begin{tabular}{lcc}\hline \hline
  Final state                               & \multicolumn{2}{c}{Parametrization form}  \\ \hline
  $\pip a^-_2$/$\pim a^+_2$                 &            & $\pm g_e$    \\
  $\kap K^{\star -}_2$/$\kam K^{\star +}_2$ & $\mp g_m$  & $\pm g_e$    \\
  $\kaz \overline{K}^{\star 0}_2$/$\kazb {K}^{\star 0}_2$
                                            & $\mp g_m$  &              \\  \hline \hline
\end{tabular}
\end{table}

\begin{table*}[bth]
\caption{\label{vpmsnfmtot}Amplitude parametrization form for decays of the $\psp$ or $\jpsi$ into $V~P$ final states. The results are based on the effective Hamiltonian in Eq.~\eref{effhmtvpsao}. With the assumptions in the last three columns, the results are the same as those due to the nonet approach.}
\begin{ruledtabular}
\begin{tabular}{lccccccccccccccc}
  States    &$g_0^{88}$&$g_m^{88}$&$g_e^{88}$&$g_0^{11}$&$g_m^{11}$&$g_e^{11}$
     &$g_0^{18}$&$g_m^{18}$&$g_e^{18}$&$g_0^{81}$&$g_m^{81}$&$g_e^{81}$
&$g_0^{88}=g_0^{11}$&$g_m^{88}=g_m^{18}=g_m^{81}$&$g_e^{88}=g_e^{18}=g_e^{81}$ \\ \hline
$\roz \piz$ &  1 &$-2/3$ &$ 1/3$ &0&0&0&0&0&0&0&0&0&  1  &$-2/3$ &$ 1/3$ \\
$\rop \pim$ &  1 &$-2/3$ &$ 1/3$ &0&0&0&0&0&0&0&0&0&  1  &$-2/3$ &$ 1/3$ \\
$\rop \pim$ &  1 &$-2/3$ &$ 1/3$ &0&0&0&0&0&0&0&0&0&  1  &$-2/3$ &$ 1/3$ \\
$\kstp\kam$ &  1 &$ 1/3$ &$ 1/3$ &0&0&0&0&0&0&0&0&0&  1  &$ 1/3$ &$ 1/3$ \\
$\kstm\kap$ &  1 &$ 1/3$ &$ 1/3$ &0&0&0&0&0&0&0&0&0&  1  &$ 1/3$ &$ 1/3$ \\
$\kstz\kazb$&  1 &$ 1/3$ &$-2/3$ &0&0&0&0&0&0&0&0&0&  1  &$ 1/3$ &$-2/3$ \\
$\kstzb\kaz$&  1 &$ 1/3$ &$-2/3$ &0&0&0&0&0&0&0&0&0&  1  &$ 1/3$ &$-2/3$ \\
$\omega^8\eta^8$&  1 &$ 2/3$ &$-1/3$ &0&0&0&0&0&0&0&0&0&  1  &$ 2/3$ &$-1/3$   \\
$\omega^8\piz$  & 0 & 0 &$1/\sqrt{3}$ &0&0&0&0&0&0&0&0&0& 0 & 0 &$1/\sqrt{3}$  \\
$\roz\eta^8$    & 0 & 0 &$1/\sqrt{3}$ &0&0&0&0&0&0&0&0&0& 0 & 0 &$1/\sqrt{3}$  \\
$\omega^1\eta^1$& 0 & 0 & 0   &1&0&0&0&0&0&0&0&0&  1  & 0  & 0   \\
$\omega^1\eta^8$&0&0&0&0&0&0&0&$-2\sqrt{2}/{3}$&$\sqrt{2}/3$&0&0&0
                                     & 0  &$-2\sqrt{2}/{3}$&$\sqrt{2}/3$   \\
$\omega^1\piz$  &0&0&0&0&0&0&0&0&$\sqrt{2/3}$&0&0&0       & 0 &0&$\sqrt{2/3}$   \\
$\omega^8\eta^1$&0&0&0&0&0&0&0&0&0&0&$-2\sqrt{2}/{3}$&$\sqrt{2}/3$
                                     & 0  &$-2\sqrt{2}/{3}$&$\sqrt{2}/3$   \\
$\roz\eta^1$    &0&0&0&0&0&0&0&0&0&0&0&$\sqrt{2/3}$       & 0 &0&$\sqrt{2/3}$   \\
\end{tabular}
\end{ruledtabular}
\end{table*}

\begin{table*}[bth]
\caption{\label{vpmsnfmmix}Amplitude parametrization form for decays of the $\psp$ or $\jpsi$ into $V~P$ final states. The mixing between octet and singlet are taken into account according to Eqs.~\eref{mixpsmsnrv} and \eref{mixvtmsnrv}. The shorthand symbols are defined as $s_{\alpha}=\sin \theta_{\alpha}$ and $c_{\alpha}=\cos \theta_{\alpha}$ $(\alpha=V,P)$. }
\begin{ruledtabular}
\begin{tabular}{ccccc}
Decay mode    &\multicolumn{4}{c}{Coupling constant} \\
$\psi \to X$  &$g_0^{88}$   &  $g_0^{11}$  &   $g_m$  &$g_e$ \\ \hline
$\roz \piz$ &  1 &0  &$-2/3$ &$ 1/3$ \\
$\rop \pim$ &  1 &0  &$-2/3$ &$ 1/3$ \\
$\rop \pim$ &  1 &0  &$-2/3$ &$ 1/3$ \\
$\kstp\kam$ &  1 &0  &$ 1/3$ &$ 1/3$ \\
$\kstm\kap$ &  1 &0  &$ 1/3$ &$ 1/3$ \\
$\kstz\kazb$&  1 &0  &$ 1/3$ &$-2/3$ \\
$\kstzb\kaz$&  1 &0  &$ 1/3$ &$-2/3$ \\
$\phi\eta$  &$c_V c_P$ &$s_V s_P$
  &$\frac{2}{3} c_V c_P+\frac{2\sqrt{2}}{3}(c_V s_P +s_V c_P)$
  &$-\frac{1}{3} c_V c_P-\frac{\sqrt{2}}{3}(c_V s_P +s_V c_P)$ \\
$\phi\etap$ &$c_V s_P$ &$-s_V c_P$
  &$\frac{2}{3} c_V s_P-\frac{2\sqrt{2}}{3}(c_V c_P -s_V s_P)$
  &$-\frac{1}{3} c_V s_P+\frac{\sqrt{2}}{3}(c_V c_P -s_V s_P)$ \\
$\omega\eta$&$s_V c_P$ &$-c_V s_P$
  &$\frac{2}{3} s_V c_P-\frac{2\sqrt{2}}{3}(c_V c_P -s_V s_P)$
  &$-\frac{1}{3} s_V c_P+\frac{\sqrt{2}}{3}(c_V c_P -s_V s_P)$ \\
$\omega\etap$&$s_V s_P$ &$c_V c_P$
  &$\frac{2}{3} s_V s_P-\frac{2\sqrt{2}}{3}(c_V s_P +s_V c_P)$
  &$-\frac{1}{3} s_V s_P+\frac{\sqrt{2}}{3}(c_V s_P +s_V c_P)$ \\
$\roz\eta$  &   0 &   0 &0 &$\sqrt{\frac{1}{3}}c_P-\sqrt{\frac{2}{3}}s_P$ \\
$\roz\etap$ &   0 &   0 &0 &$\sqrt{\frac{1}{3}}s_P+\sqrt{\frac{2}{3}}c_P$ \\
$\phi\piz$  &   0 &   0 &0 &$\sqrt{\frac{1}{3}}c_V-\sqrt{\frac{2}{3}}s_V$ \\
$\omega\piz$&   0 &   0 &0 &$\sqrt{\frac{1}{3}}s_V+\sqrt{\frac{2}{3}}c_V$ \\
\end{tabular}
\end{ruledtabular}
\end{table*}

A remark is in order here. The physical isoscalars are mixtures of the $SU(3)$ wave function $\psi_8$ and $\psi_1$
\beq
\left.\begin{array}{rcl}
f^{\prime} &=&\psi_8~\cos \theta -\psi_1~\sin \theta~,   \\
f &=&\psi_8~\sin \theta +\psi_1~\cos \theta~,
\end{array}\right.
\label{mixmsnffp}
\eeq
where $\theta$ is the nonet mixing angle and
\beq
\left.\begin{array}{rcl}
\psi_8 &=& (u\bar{u}+d\bar{d}-2s\bar{s})/\sqrt{6}~,   \\
\psi_1 &=& (u\bar{u}+d\bar{d}+s\bar{s})/\sqrt{3}~.
\end{array}\right.
\label{defofoctasig}
\eeq
These mixing relations are often rewritten to exhibit the $u\bar{u}+d\bar{d}$ and $s\bar{s}$ components which decouple for the ``ideal'' mixing angle, such that $\tan \theta_i=1/\sqrt{2}$ (or $\theta_i=35.3^{\circ}$). Defining $\alpha=\theta+54.7^{\circ}$, one obtains the physical isoscalar in the flavor basis
\beq
f^{\prime} = \frac{1}{\sqrt{2}}(u\bar{u}+d\bar{d}) \cos \alpha -s\bar{s}~\sin \alpha~,
\label{defoffpp}
\eeq
and
\beq
f = \frac{1}{\sqrt{2}} (u\bar{u}+d\bar{d}) \sin \alpha +s\bar{s}~\cos \alpha~,
\label{defoffp}
\eeq
which is the orthogonal partner of $f^{\prime}$ (replace $\alpha$ by $\alpha-90^{\circ}$). Thus for ideal mixing ($\alpha_i=90^{\circ}$),  $f^{\prime}$ becomes pure $s\bar{s}$ and the $f$ pure $u\bar{u}+d\bar{d}$.

Let's return to the ${C}$-parity transformation issue. Under the transformation of the generalized inherent ${C}$-parity, $\hat{C} [O_1 O_2]_{x} \to \xi_{x} [O_1 O_2]_{x}$, where $x=0,d,f$, that is $\xi_{0}=+1,\xi_{d}=+1,\xi_{f}=-1$. In addition, under ${C}$-parity transformation, $\hat{C} O_i \to \eta_{O_i} O_i, (i=1,2)$, synthetically,
\beq
\hat{C}~[O_1 O_2]_{x}  = \eta_{O_1} \eta_{O_2} \xi_x [O_1 O_2]_{x}~,  \\
\label{ctfmnforokt}
\eeq
At the same time for the initial state of $\psi$
\beq
\hat{C}~\psi  = \eta_{\psi} \psi~~.  \\
\label{ctfmnforpsi}
\eeq
Therefore, the term $[O_1 O_2]_{x}$ is allowed in the effective Hamiltonian as long as $\eta_{\psi}=-1=\eta_{O_1} \eta_{O_2} \xi_x$. Otherwise, the term is forbidden. With this
criterion, it is easy to figure out what kind of terms can be presented in the effective Hamiltonian for various kinds of final states, the results are summarized in Table~\ref{termalwineffh}.

By virtue of Table~\ref{termalwineffh}, two types of Hamiltonian forms exist. One type contains both $[O_1 O_2]_{0}$ and $[O_1 O_2]_{d}$ terms, while the other contains only $[O_1 O_2]_{f}$ term, that is
\beq
\Heff^{O_1 O_2} = \gz \cdot [O_1 O_2]_0
        + g_m \cdot ([O_1 O_2]_d )^3_3  + g_e \cdot ([O_1 O_2]_d )^1_1 ~,
\label{effhmtvptype}
\eeq
or
\beq
\Heff^{O_1 O_2} =  g_m \cdot ([O_1 O_2]_f )^3_3  + g_e \cdot ([O_1 O_2]_f )^1_1 ~.
\label{effhmtpptype}
\eeq

Comparing with Eqs.~\eref{effhmtvpmsn} and ~\eref{effhmtppmsn}, Eq.~\eref{effhmtvptype}
can be called the $VP$-type Hamiltonian while Eq.~\eref{effhmtpptype} the $PP$-type Hamiltonian.
For the most general case of $PP$-type Hamiltonian, the mesons of final state may belong to distinctive octets, take PT mode as an example, by virtue of Eq.~\eref{effhmtpptype}, the parametrization form is obtained and displayed in Table~\ref{ptmsnform}.

The second issue of this appendix is about another approach to derive the effective Hamiltonian.

Besides the nonet approach, the singlet component can be treated separately. Corresponding to the matrixes in Eqs.~\eref{oktvtmsn} and \eref{oktpsmsn}, the singlet matrixes are introduced as follows
\beq
{\mathbf S_V}=
\left(\begin{array}{ccc}
\omega^1/\sqrt{3} &         &     \\
       & \omega^1/\sqrt{3}  &     \\
       &          & \omega^1/\sqrt{3}
\end{array}\right)~~,
\label{sltvtmsn}
\eeq
and
\beq
{\mathbf S_P}=
\left(\begin{array}{ccc}
\eta^1/\sqrt{3} &         &     \\
       & \eta^1/\sqrt{3}  &     \\
       &          & \eta^1/\sqrt{3}
\end{array}\right)~~.
\label{oltpsmsn}
\eeq
Therefore, besides the octet-octet effective Hamiltonian, it also has the singlet-singlet effective Hamiltonian and the octet-singlet effective Hamiltonian, as listed in the following equations,
\beq
\left.\begin{array}{rcl}
\Heff^{88} &=&\gz^{88}[V P]_0 + g^{88}_m  ([V P]_d )^3_3  + g^{88}_e  ([V P]_d )^1_1 ~, \\
\Heff^{11} &=&\gz^{11}[S_V S_P]_0+g^{11}_m ([S_V S_P]_d )^3_3+g^{11}_e ([S_V S_P]_d )^1_1~, \\
\Heff^{18} &=&\gz^{18} [S_V P]_0 + g^{18}_m ([S_V P]_d )^3_3  + g^{18}_e ([S_V P]_d )^1_1 ~, \\
\Heff^{81} &=&\gz^{81} [V S_P]_0 + g^{81}_m ([V S_P]_d )^3_3  + g^{81}_e ([V S_P]_d )^1_1 ~.
\end{array}\right.
\label{effhmtvpsao}
\eeq

The calculation results are displayed in Table~\ref{vpmsnfmtot}. If the mixing between octet and singlet are taken into account according to Eqs.~\eref{mixpsmsnrv} and \eref{mixvtmsnrv}, the results are changed into Table~\ref{vpmsnfmmix}. If replacing $s_V$ and $c_V$ with $\sqrt{1/3}$ and $\sqrt{2/3}$ respectively, the Table II in Ref.~\cite{Haber} will be recovered. It is should be noted that for $g_m$, there is an overall normalized factor $-2/\sqrt{3}$ between our calculation and that of Ref.~\cite{Haber}. Furthermore, if we introduce a new angle and define $\sin \theta_{\gamma}\equiv\sqrt{1/3}$ and $\cos \theta_{\gamma}\equiv\sqrt{2/3}$, the more compact form of Table~\ref{vpmsnfmmix} can be obtained as shown in Table~\ref{vpmsnfmnew}.

\section*{Acknowledgment}
This work is supported in part by National Key Research and Development Program of China under Contracts No.~2023YFA1606003, No.~2023YFA1606000, No.~2020YFA0406302 and No.~2020YFA0406403.


\begin{thebibliography}{99}
\bibitem{bes}M.~Ablikim {\em et al.}, (BESIII Collaboration), Nucl. Instr. Meth. A {\bf 614}: 345 (2010).
\bibitem{yellow}Kuang-Ta Chao and Yi-Fang Wang, Int. J. Mod. Phys. A {\bf 24}, iii (2009)

\bibitem{Kowalski:1976mc}H.~Kowalski and T.~F.~Walsh,
Phys.\ Rev.\ D {\bf 14}, 852 (1976).
\bibitem{Clavelli:1983}L.~J.~Clavelli and G.~W.~Intemann,
Phys.\ Rev.\ D {\bf 28}, 2767 (1983).
\bibitem{Haber}H.~E.~Haber and J.~Perrier, \Journal\PRD{32}{2961}{1985}.
\bibitem{Seiden88}A.~Seiden, H.~F.-W.~Sadrozinski, and H.~E.~Haber, \Journal\PRD{38}{824}{1988}.
\bibitem{nMorisita91}N.~Morisita, I.~Kitamura and T.~Teshima,
Phys.\ Rev.\  D {\bf 44}, 175 (1991).
\bibitem{rBaldini98}R.~Baldini {\em et al.}, Phys. Lett. B {\bf 444}, 111 (1998).
\bibitem{zmy2015}K.~Zhu, X.~H.~Mo, and C.~Z.~Yuan, Int. J. Mod. Phys. A {\bf 30} (2015) 1550148
\bibitem{Baldini19}R.~B.~Ferroli {\it et al.}, Phys.\ Lett.\  B {\bf 799}, 135041 (2019).
\bibitem{moxh2022}X.~H.~Mo and J.~Y.~Zhang, Phys.\ Lett.\  B {\bf 826}, 136927 (2022).
\bibitem{moxh2024}X.~H.~Mo, P.~Wang, and J.~Y.~Zhang, Phys.\ Rev.\ D {\bf 107}, 094009 (2023).

\bibitem{Muraskin1963}M.~Muraskin and S.~L.~Glashow, Phys. Rev. {\bf 132}, 482 (1963).
\bibitem{Gupta1964a}V.~Gupta and V.~Singh, Phys. Rev. {\bf 135}, B1442 (1964).
\bibitem{Gupta1964b}V.~Gupta and V.~Singh, Phys. Rev. {\bf 136}, B782 (1964).
\bibitem{quangpham}Quang Ho-Kim and Xuan-Yem Pham, {\em Elementary Particles and Their Interactions} (Spinger-Verlag, Berlin, 1998)
\bibitem{lichtenberg}D.~B.~Lichtenberg, {\em Unitary Symmetry and Elementary Particles} (Second Edition, Academic press, New York, 1978)
\bibitem{aHosaka}A.~Hosaka and H.~Toki, {\em Quarks, Baryons and Chiral Symmetry}'' (World Scientific Publishing Co.Pte.Ltd., Singapore, 2001).

\bibitem{pdg2020}P.~A.~Zyla {\em et al.} (PDG Collaboration), Porg. Theor. Exp. Phys. {\bf 2020}, 083C01 (2020).

\bibitem{wangp03hepnp}P.~Wang, C.~Z.~Yuan, and X.~H.~Mo, High Energy Phys. Nucl. Phys. 27, 465 (2003).
\bibitem{rudaz}S.~Rudaz, \Journal\PRD{14}{298}{1976}.
\bibitem{wymcgam}P.~Wang, C.~Z.~Yuan, X.~H.~Mo, and D.~H.~Zhang,
\Journal\PLB{593}{89}{2004}.
\bibitem{Wang:2005sk}P.~Wang, X.~H.~Mo, and C.~Z.~Yuan,
Int.\ J.\ Mod.\ Phys.\  A {\bf 21}, 5163 (2006). 
\bibitem{wymphase}P.~Wang, C.~Z.~Yuan, and X.~H.~Mo,
\Journal\PRD{69}{057502}{2004}.
\bibitem{wymogpiapp}P.~Wang, X.~H.~Mo, and C.~Z.~Yuan,
\Journal\PLB{557}{192}{2003}.
\bibitem{rad.1}{E.~A.~Kuraev and V.~S.~Fadin, Yad. Fiz. {\bf 41}
       (1985) 733 [Sov. J. Nucl. Phys. {\bf 41} (1985) 466].}
\bibitem{rad.2}G.~Altarelli and G.~Martinelli, CERN {\bf 86-02}, 47 (1986);
        O.~Nicrosini and L.~Trentadue, Phys. Lett. B {\bf 196},  551 (1987).
\bibitem{rad.3}F.~A.~Berends, G.~Burgers and W.~L.~Neerven,
        Nucl. Phys.~B{\bf 297}, 429 (1988); {\it ibid.} {\bf 304}, 921 (1988).
\bibitem{Tsai}Y.~S.~Tsai, SLAC-PUB-3129 (1983).
\bibitem{Luth}P.~Alexander {\em et al.}, \Journal\NPB{320}{45}{1989}.

\bibitem{besscan95}J.~Z.~Bai {\em et al.} (BES Collaboration), \Journal\PLB{355}{374}{1995}.
\bibitem{besscan02}J.~Z.~Bai {\em et al.} (BES Collaboration), \Journal\PLB{550}{24}{2002}.

\bibitem{awChao}A.W.~Chao, {\em Lectures on Accelerator Physics} (World Scientific Publishing Co.Pte.Ltd., Singapore, 2020).

\bibitem{YELLOWBOOK} CLEO-c/CESR-c Taskforces \& CLEO-c Collaboration,
 Cornell University LEPP Report No. CLNS~01/1742 (2001) (unpublished).
\bibitem{wymhepnp}P.~Wang, C.~Z.~Yuan, and X.~H.~Mo,
\Journal\HEPNP{27}{465}{2003}.

\bibitem{tnpsp2004}X.~H.~Mo {\em et al.}, High Energy Phys. Nucl. Phys. 28, 455 (2004).
\bibitem{lum2004}S.~P.~Chi, X.~H.~Mo, and Y.~S.~Zhu, High Energy Phys. Nucl. Phys. 28, 1135 (2004).
\bibitem{tnpsp2018}M.~Ablikim {\em et al.} (BESIII Collaboration), Chin. Phys. C {\bf 42}, 023001 (2018).
\bibitem{besretp2017}M.~Ablikim {\em et al.} (BESIII Collaboration),
Phys.\ Rev.\ D {\bf 96}, 112012 (2017)
\bibitem{cleovp2005}N.~E.~Adam {\em et al.} (CLEO Collaboration),
\Journal\PRL{94}{012005}{2005}.  
\bibitem{tnjps2012}M.~Ablikim {\em et al.} (BESIII Collaboration), CPC (HEP {\&} NP), 36, 915 (2012).
\bibitem{bes3pi2012}M.~Ablikim {\em et al.} (BESIII Collaboration),
\Journal\PLB{710}{594}{2012}.
\bibitem{tnjps2017}M.~Ablikim {\em et al.} (BESIII Collaboration), Chin. Phys. C {\bf 41}, 013001 (2017).
\bibitem{tnjps2003}S.~S.~Fang {\em et al.}, High Energy Phys. Nucl. Phys. 27, 277 (2003).
\bibitem{besjopaet06}M.~Ablikim {\em et al.} (BESII Collaboration),
Phys.\ Rev.\ D {\bf 73}, 052007 (2006)
\bibitem{JJousset90jvp}J.~Jousset {\it et al.} (DM2 Collaboration),
Phys.\ Rev.\ D {\bf 41}, 1389 (1990).

\bibitem{mk3jvp85}D.~Coffman {\em et al.}, (Mark III Collaboration),
                Phys. Rev. D{\bf 32}, 2883 (1985).
\bibitem{mk3jvp88}D.~Coffman {\em et al.}, (Mark III Collaboration),
                Phys. Rev. D{\bf 38}, 2695 (1988).

\bibitem{a00}M.~Suzuki, Phys. Rev. D{\bf 60}, 051501 (1999).
\bibitem{LopezCastro:1994xw}G.~L\'{o}pez Castro, J.~L.~Lucio M., and J.~Pestieau,
AIP Conference Proceedings 342, 441 (1995); arXiv:hep-ph/9902300.
\bibitem{wymppdk}C.~Z.~Yuan, P.~Wang, and X.~H.~Mo, \Journal\PLB{567}{73}{2003}.
\bibitem{bes2klks04}J.~Z.~Bai {\em et al.} (BES Collaboration), Phys. Rev. D{\bf 69}, 012003 (2004).
\bibitem{a11}L.~K\"{o}pke and N.~Wermes, Phys. Rep. {\bf 174}, 67 (1989).

\bibitem{Wang:2003zx}P.~Wang, C.~Z.~Yuan, and X.~H.~Mo,
Phys.\ Lett.\ B{\bf 574}, 41 (2003).
\bibitem{gerard}J.~M.~G\'{e}rard and J.~Weyers,
\Journal\PLB{462}{324}{1999}.

\bibitem{Franklin83}M.~E.~B.~Franklin {\em et al.}, Phys. Rev. Lett. {\bf 51} (1983) 963.
\bibitem{wBartel76}W.~Bartel {\em et al.}, Phys. Lett. {\bf 64}B, 483 (1976).

\bibitem{besrp2005}M.~Ablikim {\em et al.} (BESII Collaboration),
\Journal\PLB{619}{247}{2005}.
\bibitem{beskstk2005}M.~Ablikim {\em et al.} (BESII Collaboration),
\Journal\PLB{614}{37}{2005}.
\bibitem{besfw2004}M.~Ablikim {\em et al.} (BESII Collaboration),
Phys.\ Rev.\ D {\bf 70}, 112003 (2004)
\bibitem{besvps2012}M.~Ablikim {\em et al.} (BESIII Collaboration),
Phys.\ Rev.\ D {\bf 87}, 072011 (2012)

\bibitem{besfet2019}M.~Ablikim {\em et al.} (BESIII Collaboration),
Phys.\ Rev.\ D {\bf 100}, 092003 (2019)
\bibitem{besret2004}M.~Ablikim {\em et al.} (BESII Collaboration),
Phys.\ Rev.\ D {\bf 70}, 112007 (2004)

\bibitem{moxh2010prd}X.~H.~Mo, C.~Z.~Yuan, and P.~Wang, \Journal\PRD{82}{077501}{2010}.
\bibitem{yuancz2010ijmpa}C.~Z.~Yuan, X.~H.~Mo, and P.~Wang, Int. J. Mod. Phys. A, 25: 5963 (2010)

\bibitem{yuancz2011cpc}C.~Z.~Yuan, X.~H.~Mo, and P.~Wang, CPC (HEP{\&}NP), {\bf 35}, 543 (2011).
\bibitem{zhuk2011ijmpa}K.~Zhu, X.~H.~Mo, C.~Z.~Yuan, and P.~Wang, Int. J. Mod. Phys. A 26, 4511 (2011).
\bibitem{hanx2018cpc}X.~Han and C.~P.~Shen, Chin. Phys. C {\bf 42}, 043001 (2018).
\bibitem{baiyu2019prd}Y.~Bai and D.~Y.~Chen, Phys. Rev. D {\bf 99}, 072007 (2019).

\bibitem{besj3pi04}M.~Ablikim {\em et al.} (BES Collaboration),
Phys.\ Rev.\ D {\bf 70}, 012005 (2004)
\bibitem{wBartelg1976}W.~Bartelg {\it et al.} (DESY-Heidelberg Collaboration),
Phys.\ Lett.\  {\bf 64} B, 483 (1976).
\bibitem{besjfpaet05}M.~Ablikim {\em et al.} (BESII Collaboration),
Phys.\ Rev.\ D {\bf 71}, 032003 (2005)
\bibitem{bes16fetp}M.~Ablikim {\em et al.} (BESIII Collaboration),
Phys.\ Rev.\ D {\bf 93}, 072008 (2016)
\bibitem{bes15fpi}M.~Ablikim {\em et al.} (BESIII Collaboration),
Phys.\ Rev.\ D {\bf 91}, 112001 (2015)
\bibitem{bJeanMarie76}B.~Jean-Marie {\em et al.} (MARKI Collaboration), Phys. Rev. Lett. {\bf 36} (1976) 291.
\bibitem{gAlexander1978}G.~Alexander {\it et al.} (PLUTO Collaboration),
Phys.\ Lett.\  {\bf 72} B, 493 (1978).
\bibitem{Brandelik:1978}R.~Brandelik {\it et al.} (DASP Collaboration),
Phys.\ Lett.\  {\bf 74} B, 292 (1978).
\bibitem{wBraunschweig1976}W.~Braunschweig {\it et al.} (DASP Collaboration),
Phys.\ Lett.\  {\bf 63} B, 487 (1976).
\bibitem{fVannucci77}F.~Vannucci {\em et al.} (MARKI Collaboration),
Phys.\ Rev.\ D {\bf 15}, 1814 (1977)

\bibitem{bes3pi96}J.~Z.~Bai {\it et al.} (BES Collaboration),
Phys.\ Rev.\  D {\bf 54}, 1221 (1996). 
\bibitem{bes09ksk}M.~Ablikim {\em et al.} (BESIII Collaboration),
Phys.\ Rev.\ D {\bf 100}, 032004 (2019)
\bibitem{aAubert3pi04}B.~Aubert {\em et al.} (BaBar Collaboration), Phys. Rev. D{\bf 70} (2004) 072004.
\bibitem{aAubert07oget}B.~Aubert {\em et al.} (BaBar Collaboration), Phys. Rev. D{\bf 73} (2006) 052003.


\bibitem{IGHuges2010}Ifan G.~Hughes and Thomas P.~A.~Hase, {\em Measurements and their Uncertainties} (Oxford University Press 2010)
\bibitem{AGFrodeson19790}A.~G.~Frodeson, O.~Skjeggestad, and H.~T\o fte, ``{\em Pobability and statitics in particle physics}'' (Universitetsforlaget, Bergen-Oslo-Troms\o , 1979)

\bibitem{pdg2022}R.~L.~Workman {\em et al.} (Particle Data Group), Prog. Theor. Exp. Phys. 2022, 083C01 (2022)
\bibitem{ozi}S. Okubo, \Journal\PL{5}{165}{1963};
G.~Zweig, CERN-preprint CERN-TH-401, 402, 412 (1964);
Iizuka, \Journal\PTPS{37-38}{21}{1996}.
\bibitem{appelquist}T.~Appelquist and H.~D.~Politzer,
Phys. Rev. Lett. {\bf 34}, 43 (1975); A.~De R\'{u}jula and
S.~L.~Glashow, Phys. Rev. Lett. {\bf 34}, 46 (1975).

\bibitem{moxh07rprv}X.~H.~Mo, C.~Z.~Yuan, and P.~Wang, High Energy Phys. Nucl. Phys. 31, 686 (2007).

\bibitem{brodsky81}S.~J.~Brodsky and G.~P.~Lepage, \Journal\PRD{24}{2848}{1981}.
\bibitem{freund}P.~G.~O.~Freund and Y.~Nambu, \Journal\PRL{34}{1645}{1975}.
Iizuka, \Journal\PTPS{37-38}{21}{1996}.
\bibitem{houws}W.~S.~Hou and A.~Soni, \Journal\PRL{50}{569}{1983}.
\bibitem{houwsa}W.~S.~Hou and A.~Soni, \Journal\PRD{29}{101}{1984}.
\bibitem{brodsky87}S.~J.~Brodsky, G.~P.~Lepage, and S.~F.~Tuan, \Journal\PRL{59}{621}{1987}.
\bibitem{chan}C.~T.~Chan and W.~S.~Hou, \Journal\NPA{675}{367c}{2000}.
\bibitem{houwsb}W.~S.~Hou, \Journal\PRD{55}{6952}{1997}.
\bibitem{anselm}M.~Anselmino, M.~Genovese, and E.~Predazzi, Phys. Rev. D {\bf 44}, 1597 (1991);
M.~Anselmino, M.~Genovese, and D.~E.~Kharzeev, ibid. {\bf 50}, 595 (1994).
\bibitem{chaichian}M.~Chaichian and N.~A.~T\"{o}rnqvist, \Journal\PLB{323}{75}{1989}.
\bibitem{brokar}S.~J.~Brodsky and M.~Karliner, \Journal\PRL{78}{4682}{1997}.

\bibitem{chen}Y.~Q.~Chen and E.~Braaten, \Journal\PRL{80}{5060}{1998}.
\bibitem{karl}G.~Karl and W.~Roberts,
\Journal\PLB{144}{263}{1984}. 
\bibitem{karla}G.~Karl and S-F.~Tuan, \Journal\PRD{34}{1692}{1986}.
\bibitem{pinsky}S.~S.~Pinsky, \Journal\PLB{236}{479}{1990}.

\bibitem{artoisenet}P.~Artoisenet, J.~M.~G\'{e}rard, and J.~Weyers,
\Journal\PLB{628}{211}{2005}.
\bibitem{Chernyak}V.~L.~Chernyak and A.~R.~Zhitnitsky,
\Journal\PRP{112}{173}{1984}.
\bibitem{rosnersd}J.~L.~Rosner, Phys. Rev. D {\bf 64}, 094002 (2001).

\bibitem{lixq}X.~Q.~Li, D.~V.~Bugg, and B.~S.~Zou, \Journal\PRD{55}{1421}{1997}.
\bibitem{suzukia}M.~Suzuki,\Journal\PRD{57}{5171}{1998}.
\bibitem{suzukib}M.~Suzuki, \Journal\PRD{60}{051501}{1999}.
\bibitem{suzukic}M.~Suzuki, \Journal\PRD{63}{054021}{2001}.
\bibitem{majp}J.~P.~Ma, \Journal\PRD{65}{097506}{2002}.
\bibitem{clavelli}L.~J.~Clavelli and G.~W.~Intemann,
\Journal\PRD{28}{2767}{1983}.
\bibitem{feldman}T.~Feldmann and P.~Kroll, \Journal\PRD{62}{074006}{2000}.
\bibitem{hoyer}P.~Hoyer and S.~Peign\'{e}, \Journal\PRD{61}{031501}{2000}.

\bibitem{psippkskl}P.~Wang, X.~H.~Mo, and C.~Z.~Yuan, \Journal\PRD{70}{077505}{2004}.

\end{thebibliography}
\end{document}